\documentstyle[amssymb,12pt,epsf]{article}

   \newlength{\absize}
\begin{document}
  \newcommand\beq{\begin{equation}}
  \newcommand\eeq{\end{equation}}
\renewcommand{\title}[1]{    \begin{center}
      \LARGE #1
    \end{center}\par} 
\renewcommand{\author}[1]{    \vspace{2ex}
    {\normalsize
     \begin{center}
       \setlength{\baselineskip}{3ex} #1 \par
     \end{center}}} \renewcommand{\thanks}[1]{\footnote{#1}} 
\renewcommand{\abstract}[1]{    \vspace{2ex}
    \normalsize
    \begin{center}
      \centerline{\bf Abstract}\par
      \vspace{2ex}
      \parbox{\absize}{#1\setlength{\baselineskip}{2.5ex}\par}
    \end{center}} 
\begin{titlepage}
  
\vspace{3mm}

  \title{The structure function
of semi-inclusive heavy flavour decays in field theory}

  \vskip 0.1cm

\begin{center}
{\large Ugo Aglietti$\; {}^{a}$\footnote{\sl e-mail: 
ugo.aglietti@cern.ch} and
Giulia Ricciardi$\; {}^{b}$\footnote{\sl e-mail:
giulia.ricciardi@na.infn.it}}

\vspace{3mm}

$^a${\sl
CERN-TH, Geneva, Switzerland, 
and I.N.F.N., Sezione di Roma, Italy.}

\vspace{2mm}

$^b${\sl
Dipartimento di Scienze Fisiche,
Universit\'a di Napoli {\sl ``Federico II''}\\
and I.N.F.N., Sezione di Napoli, Italy.}\\

\end{center}

\vspace{4mm}

\abstract{
We consider the decay of 
a heavy flavour into an inclusive hadronic state $X$ of
invariant mass $m_X$ small with respect to its energy $E_X$, $m_X \ll E_X$. 
The electron spectrum and the hadronic mass distribution in semileptonic 
$b\rightarrow u$
decays, or the photon spectrum in $b\rightarrow s\gamma$ decays, all require,
close to their endpoints, a control over this region. 
This region is affected both by non-perturbative phenomena related to the
Fermi motion of the heavy quark and by perturbative 
soft gluon radiation in the final state (Sudakov form factor). 
Fermi motion can be described 
by the shape function $f (m_*)$, 
which represents the distribution of the effective mass $m_*$ 
of the heavy quark at disintegration time.
  We perform a factorization with a simple technique 
 in order to  consistently separate
perturbative from non-perturbative effects. 
We find that the shape function,
contrary to naive expectations,
 is not a physical distribution, as it is affected 
by substantial regularization scheme effects, 
controlling even the leading, double-logarithmic term.
It factorizes, however,
the bulk of non-perturbative effects in lattice-like
regularizations.
Some non-perturbative effects are present in the coefficient function 
even at leading twist, but they are expected 
to be suppressed on physical grounds. Finally, we clarify a controversial
factor of 2 in the evolution kernel of the shape function.}

\noindent



Los Alamos-no: hep-ph/0003146 

Report-no: CERN-TH/2000-071,DSF-T/2000-7

\end{titlepage}

\setcounter{footnote}{0}

\newpage

\section{ Introduction}

Nowadays there are many facilities that allow an accurate experimental study
of heavy flavour decays. It is therefore becoming more and more important to
improve the accuracy and the reliability of the theoretical calculations. In
this paper, we study the properties of the decays of heavy flavour hadrons
into inclusive hadronic states $X$ \ ``tight'' in mass, i.e. with an
invariant mass $m_{X}$ small with respect to the energy $E_{X}$: 
\begin{equation}
m_{X}\ll E_{X}.  \label{prima}
\end{equation}
More specifically, we consider the situation where 
\begin{equation}
m_{X}^{2}\sim O\left( E_{X}\,\Lambda _{QCD}\right) ,  \label{seconda}
\end{equation}
so that 
\begin{equation}
\frac{m_{X}^{2}}{E_{X}^{2}}\sim O\left( \frac{\Lambda _{QCD}}{E_{X}}\right)
\ll 1\qquad \left( E_{X}\gg \Lambda _{QCD}\right) .  \label{smallra}
\end{equation}
A formal definition of kinematics (\ref{seconda}) is the limit, in the heavy
quark rest frame: 
\begin{eqnarray*}
E_{X} &\rightarrow &\infty , \\
m_{X}^{2} &\rightarrow &\infty 
\end{eqnarray*}
with 
\begin{equation}
\frac{m_{X}^{2}}{E_{X}}={\rm const.}  \label{one}
\end{equation}
The divergence of $m_{X}^{2}$ - even though slower than the one of $E_{X}^{2}
$ - implies that the final hadronic state can be replaced with a partonic
one, i.e. that the use of perturbation theory is fully justified. Heavy
flavour decays are characterized by three mass or energy scales: the mass of
the heavy flavour $m_{Q}$, the energy $E_{X}$, and the invariant mass $m_{X}$
of the final hadronic state. Limit (\ref{one}) implies also the limit of
infinite mass for the heavy flavour $Q$: 
\begin{equation}
m_{Q}\rightarrow \infty ,  \label{hqet}
\end{equation}
since $m_{Q}\geq E_{X}$. Another consequence of (\ref{one})\ is that 
\begin{equation}
\frac{m_{X}^{2}}{E_{X}^{2}}\,\rightarrow \,0,  \label{Sudakov}
\end{equation}
i.e. we are in the so-called threshold region\footnote{%
The converse is not true: limit (\ref{Sudakov}) does not imply limit (\ref
{one}) (we thank G. Veneziano for pointing this out to us).}.

The study of these processes has both a theoretical and a phenomenological
interest. On the theoretical side, in heavy decays the infrared perturbative
structure of gauge theories - the Sudakov form factor \cite
{Sudakov1,Sudakov2} - enters in a rather ``pure'' form, owing to the absence
of initial state mass singularities. On the phenomenological side, the
computation of many relevant distributions requires a good theoretical
control over the region (\ref{prima}). As examples, let us quote the
electron spectrum $d\Gamma /dE_{e}$ close to the endpoint $E_{e}\lesssim
m_{B}/2$ \cite{accmm} and the hadron mass distribution $d\Gamma /dm_{X}$ at
small $m_{X}$ \cite{uraltsev} in semileptonic $b\rightarrow u$ decays, such
as 
\begin{equation}
B\rightarrow X_{u}+e+\nu ,  \label{semilep}
\end{equation}
or the photon energy distribution $d\Gamma /dE_{\gamma }$ close to the
endpoint $E_{\gamma }\lesssim m_{B}/2$ in $b\rightarrow s\gamma $ decays.
For the electron or photon spectrum, the region (\ref{prima}) is involved
because the requirement of a large energy of \ the lepton or of the photon
pushes down to zero the mass of the recoiling hadronic system. As is well
known, the above mentioned distributions in (\ref{semilep}) allow an
inclusive determination of\ the CKM-matrix element $|V_{ub}|$ \cite{volkov},
while a large photon energy in the rare decay, $E_{\gamma }\gtrsim 2.1$ GeV
is required to cut experimental backgrounds.

In general, the dynamics in region (\ref{prima}) is rather intricate as it
involves an interplay of non-perturbative and perturbative contributions.
These are related to the Fermi motion of the heavy quark inside the hadron
and to the Sudakov suppression in the threshold region (\ref{Sudakov}),
respectively. Even though these two effects are physically distinguishable
and are treated as independent in various models \cite{accmm}, they are
ultimately both described by the same quantum field theory, QCD. Therefore
the problem arises of describing them consistently, i.e. without double
countings, inconsistencies, etc. Our idea is to subtract from the hadronic
tensor encoding $all$ QCD dynamics, 
\begin{equation}
W_{\mu \nu }\ \equiv \sum_{X}\,\langle \,H_{Q}|J_{\nu }^{+}|X\rangle
\,\langle X|J_{\mu }|H_{Q}\rangle \,\delta ^{4}(p_{B}-q-p_{X}),
\label{Wmunu}
\end{equation}
each of the perturbative components - associated with the Sudakov form
factor and with other short-distance corrections - to end up with an
explicit representation of the non-perturbative component. In eq.~(\ref
{Wmunu}), we have defined 
\begin{equation}
J_{\mu }(x)\equiv \overline{q}(x)\Gamma _{\mu }Q(x),
\end{equation}
where $\Gamma _{\mu }$ is a matrix in Dirac algebra\footnote{%
For the left-handed currents of the Standard Model, $\Gamma _{\mu }=\gamma
_{\mu }\left( 1-\gamma _{5}\right) $.}, $p_{X}$ is the momentum of the final
hadronic jet, and $H_{Q}$ is a hadron containing the heavy quark $Q$. The
non-perturbative component is identified with an ultraviolet (UV)
regularized expression for the structure function, or shape function, in the effective theory. The
shape function $f(k_{+})$ has been introduced using the Operator Product
Expansion $(OPE)$ and can be defined as \cite{generalshape} 
\begin{equation}
f(k_{+})\equiv \langle H_{Q}\mid h_{v}^{\dagger }\,\delta
(k_{+}-iD_{+})\,h_{v}\mid H_{Q}\rangle ,  \label{defshape}
\end{equation}
where $h_{v}$ is a field in the Heavy Quark Effective Theory (HQET) with
4-velocity $v$; $D_{+}$ is the plus component of the covariant derivative,
i.e. $D_{+}\equiv D^{0}+D^{3}$. The shape function represents the
probability that the heavy quark has a momentum $m_{B}v+k^{\prime }$ \ with
a given plus component $k_{+}^{\prime }=k_{+}$. This function can also be
interpreted (see section~\ref{variable_mass}) as the probability that $Q$
has an effective mass 
\begin{equation}
m_{\ast }=m_{B}+k_{+}  \label{massvar}
\end{equation}
at disintegration time. The renormalization properties of the shape function
have also been analysed~\cite{renshape,kg,mn,noi}. Because of UV divergences
affecting its matrix elements, $f(k_{+})$ needs to be renormalized and it
consequently acquires a dependence on the renormalization point $\mu $: $%
f(k_{+};\mu )$. The non-perturbative information about Fermi motion enters
in this framework as the initial value for the $\mu $-evolution. The shape
function can be extracted from a reference process and used to predict other
processes, analogously to the parton distribution functions in usual hard
processes such as Deep Inelastic Scattering (DIS) or Drell--Yan \cite
{altmart}. In principle, it can also be computed with a non-perturbative
technique, for example lattice QCD \cite{lattice}.

Our approach aims at a deeper understanding of perturbative and
non-perturbative effects with respect to the standard OPE in dimensional
regularization (DR). We compare different regularization schemes and find
that the factorization procedure is substantially scheme-dependent. By that,
we mean a much stronger scheme dependence than the usual one, typically
related to the finite part of one-loop amplitudes, corresponding in DR to a
replacement of the form $1/\epsilon \rightarrow 1/\epsilon +$const. The
shape function, in contrast to naive expectations, is not a physical
distribution, but it is affected by regularization scheme effects even at
the leading, double-logarithmic level. We show, however, that it factorizes
most of the non perturbative effects in a class of regularization schemes.

This paper is devoted to a wide audience, i.e. not only to perturbative QCD
experts, but also to phenomenologists who are interested in the field
theoretic aspects of this area of $B$ physics, as well as to lattice-QCD
physicists who may wonder about the possibility of simulating the shape
function. We have therefore tried to give a plain presentation of our
method, together with a self-consistent description of the known results to
be found in the literature.

In section~\ref{intro_plain} we give a simple introduction to the physics of
semi-inclusive heavy flavour decays. In section~\ref{strategy} we present
our strategy, based on factorization, in order to consistently combine
perturbative and non-perturbative contributions and to arrive at a formal
definition of the shape function in field theory; we outline the main steps
and the relevant issues. In section~\ref{opesection} we review the standard $%
OPE$ derivation of the shape function in the effective theory; this section
can be skipped by the experienced reader. In section~\ref
{factorization_quantum} we return to the strategy outlined in section~\ref
{opesection} and apply our factorization procedure in the quantum theory to
a specific class of loop corrections. Our technique is completely general,
but we believe that it is better illustrated by treating in detail a simple
computation, which illustrates most of the general features. In section \ref
{nuova} we discuss factorization in the framework of the effective theory on
the light-cone, the so called Large Energy Effective Theory (LEET), which is
the relevant effective theory for these processes at low energy. In section~%
\ref{effective_theory} we describe the properties of the shape function in
the effective theory in various regularizations and its evolution with the
UV cutoff or renormalization scale. We also discuss our results on
factorization and clarify a controversial factor of 2 in the evolution
kernel of the shape function. Section~\ref{sec_conclusion} contains the
conclusions.

\section{Physics of semi-inclusive 
heavy flavour decays}

\label{intro_plain}

Let us begin by discussing Fermi motion. This phenomenon, originally
discovered in nuclear physics, is classically described as a small
oscillatory motion of the heavy quark inside the hadron, due to the
interaction with the valence quark; in the quantum theory it is also the 
{\it virtuality} of the heavy flavour that matters. Generally, as the mass
of the heavy flavour becomes large, i.e. as we take the limit (\ref{hqet}),
we expect that the heavy particle decouples from the light degrees of
freedom and becomes ``frozen'' with respect to strong interactions. That is
indeed true in the ``bulk'' of the phase space of the decay products, but it
is untrue close to kinematical boundaries, as in region (\ref{prima}). This
is because a kinematical amplification effect occurs, according to which a
small virtuality of the heavy flavour in the initial state produces
relatively large variations of the fragmentation mass in the final state. To
see how this works in detail, let us begin with a picture of the initial
bound state. We assume that the momentum exchanges $r_{\mu }$ between the
heavy flavour and the light degrees of freedom are of the order of the
hadronic scale, 
\begin{equation}
|r_{\mu }|\sim O\left( \Lambda _{QCD}\right) ,  \label{typical}
\end{equation}
as we take the infinite mass limit (\ref{hqet}). In other words, we assume
that the momentum transfer does not scale with the heavy mass but remains
essentially constant\footnote{%
We must specify that we consider an initial hadron containing a {\it single}
heavy quark: hadrons containing {\it more} than one heavy quark, such as for
example quarkonium states, need a different theoretical treatment \cite
{matteo}.}. This assumption, which is rather reasonable from a physical
viewpoint, is at the basis of the application of the HQET~\cite{hqet}. Let
us discuss for example the decay (\ref{semilep}). The initial meson has
momentum 
\begin{equation}
p_{B}=m_{B}v,
\end{equation}
where $v$ is the 4-velocity, which we can take at rest without any loss of
generality: $v^{\mu }=(1;0,0,0)$. The final hadronic state $X$ \ has a
momentum 
\begin{equation}
Q=m_{B}v-q  \label{Qbig}
\end{equation}
and invariant mass 
\begin{equation}
m_{X}^{2}\equiv Q^{2}.
\end{equation}
In eq.~(\ref{Qbig}) $q_{\mu }$ is the momentum of the virtual $W$ or,
equivalently, of the leptonic pair. We isolate in the decay a hard
subprocess consisting in the fragmentation of the heavy quark. If the
valence quark - in general the light degrees of freedom in the hadron - have
momentum $-k^{\prime }$, the heavy quark has a momentum\footnote{%
For the appearance of $m_{B}$ instead of $m_{b}$, see footnote in section 
\ref{opesection_1}.} 
\begin{equation}
p_{Q}=m_{B}v+k^{\prime }  \label{momhq}
\end{equation}
and a virtuality 
\begin{equation}
p_{Q}^{2}-m_{B}^{2}=2m_{B}v\cdot k^{\prime }+k^{\prime 2}\neq 0\quad \left( 
{\rm in\,\,general}\right) .
\end{equation}
The final invariant mass of the hard subprocess, i.e. the fragmentation
mass, is 
\begin{equation}
\widehat{m}_{X}^{2}\equiv \left( p_{Q}-q\right) ^{2}=\left( Q+k^{\prime
}\right) ^{2}=m_{X}^{2}+2Q\cdot k^{\prime }+k^{\prime 2}\simeq
m_{X}^{2}+2Q\cdot k^{\prime }  \label{fragmass}
\end{equation}
and this is the mass that controls the kinematic of the hard subprocess,
i.e. the Sudakov form factor (the difference between $m_{X}^{2}$ and $%
\widehat{m}_{X}^{2}$ is that we do not include in the latter the momentum of
the valence quark). The term $k^{\prime 2}$ has been neglected in the last
member of eq.~(\ref{fragmass}) because it is small, as gluon exchanges are
soft according to the assumption (\ref{typical}). We take the motion of the
final $up$ quark in the $-z$ direction, so that the vector $Q$ has large
zero and third components, both of order $E_{X},$ and a small square; we
have therefore for the average in the meson state: 
\begin{equation}
\langle Q\cdot k^{\prime }\rangle =Q\cdot \langle k^{\prime }\rangle \sim
O\left( E_{X}\,\Lambda _{QCD}\right) .
\end{equation}
A fluctuation in the heavy quark momentum of order $\Lambda _{QCD}$ in the
initial state produces a variation of the final invariant mass of the hard
subprocess of order 
\begin{equation}
\delta \widehat{m}_{X}^{2}\sim O\left( \Lambda _{QCD}\,E_{X}\right) .
\label{ampl}
\end{equation}
An amplification by a factor $E_{X}$ has occurred, as anticipated. The
fluctuation (\ref{ampl}) is of the order of (\ref{seconda}) and so it must
be taken into account.

We will discuss the shape function at length in sections~\ref{opesection}
and~\ref{effective_theory}, but let us introduce now some of its more
important properties. If we consider a heavy quark with the given off-shell
momentum (\ref{momhq}), we find for the shape function\footnote{%
The final state consists of a massless on-shell quark at the tree level.} 
\begin{equation}
f(k_{+})^{part}=\delta \left( k_{+}-k_{+}^{\prime }\right) +O\left( \alpha
_{S}\right),  \label{basres}
\end{equation}
where 
\begin{equation}
k_{+}\equiv -\frac{m_{X}^{2}}{2E_{X}}.  \label{defkplus}
\end{equation}
Selecting the hadronic final state, i.e. $k_{+}$, we select the light-cone
virtuality $k_{+}^{\prime }=k_{+}$ of the heavy flavour which participates
in the decay. After inclusion of the radiative corrections, we find that in
general $k_{+}^{\prime }\geq k_{+}$~\footnote{%
To obtain the hadronic shape function, the ``elementary'' or ``partonic''
shape function in eq.~(\ref{basres}) has to be convoluted with the
distribution $\varphi _{0}\left( k_{+}^{\prime }\right) $ of the primordial
light-cone virtuality $k_{+}^{\prime }$ of the heavy quark inside the hadron.%
}. Equation~(\ref{basres}) is analogous to the relation between the Bjorken
variable $x_{B}\equiv -q^{2}/\left( 2p\cdot q\right) $ ($p$ is the momentum
of the hadron and $q$ that of the space-like photon) and the momentum
fraction $x$ of partons in the naive parton model, where we have 
\begin{equation}
q\left( x\right) ^{part}=\delta \left( x-x_{B}\right) +O\left( \alpha
_{S}\right) .
\end{equation}
In this case, as is well known, by selecting final state kinematics, i.e. $%
x_{B}$, one selects the momentum fraction $x=x_{B}$ of the partons that
participate in the hard scattering. Just as in the heavy flavour decay,
radiative corrections lead to a softening of the above condition in $x\geq
x_{B}$, due to the emission of collinear partons.

We note that even with the amplification effect (\ref{ampl}), Fermi motion
effects are irrelevant in most of the phase space, where typical values for
the final hadron mass are 
\begin{equation}
\widehat{m}_{X}^{2}\sim O(E_{X}^{2}).  \label{bulk}
\end{equation}
This is in agreement with physical intuition.

As will be proved in section~\ref{variable_mass}, the shape function can be
interpreted as the distribution of a {\it variable} mass. The virtuality of
the heavy flavour \ can be represented by a shift of its mass, $%
m_{b}\rightarrow m_{\ast }$. In other words, an off-shell particle with a
given mass, i.e. with the momentum (\ref{momhq}), can be replaced by an
on-shell particle with a variable, virtuality dependent, mass, i.e. with a
momentum $m_{\ast }\left( k_{+}\right) v$. The physical distribution is
obtained by convoluting the distribution of an isolated quark of mass $%
m_{\ast }$ with the probability distribution for such a mass (see eq.~(\ref
{funda})): this is the basis of the factorization theory for the
semi-inclusive heavy flavour decays.

Fermi motion is a non-perturbative effect in QCD because it involves low
momentum transfers to the heavy flavour (cf. eq.~(\ref{typical})), at which
the coupling is large; it does however occur also in QED bound states, where
it can be treated with perturbation theory~\footnote{%
Consider for instance an atom composed of a $\mu $ and an $e$, decaying by $%
\mu $ fragmentation.}.

The second phenomenon relevant in region (\ref{prima}) is related to soft
gluon emission and it is of a perturbative nature - it is a case of the
Sudakov form factor in QCD \cite{webber}. The $up$ quark emitted by the
fragmentation of the heavy flavour with a large virtuality - of the order of
the final hadronic energy $E_{X}$ - evolves in the final state, emitting
soft and collinear partons, either real or virtual. Since the final state is
selected to have a small invariant mass (cf. eq.~(\ref{Sudakov})), real
radiation is inhibited with respect to the virtual one. That means that
infrared (IR) singularities coming from real and virtual diagrams still
cancel, but leave a large residual effect in the form of large logarithms%
\footnote{%
The plus-distribution is defined as 
\begin{equation}
\left( \frac{\log x}{x}\right) _{+}= \theta(x)\frac{\log x}{x}-\delta \left(
x\right) \int_{0}^{1}\frac{\log y}{y}dy.
\end{equation}
}: 
\begin{equation}
\alpha _{S}\left( \frac{\log m_{X}^{2}/E_{X}^{2}}{m_{X}^{2}/E_{X}^{2}}%
\right) _{+}.  \label{largelog}
\end{equation}
Schematically, the rate for final states with an invariant mass $m_{X}^{2}$
has double-logarithmic contributions at order $\alpha _{S}$, of the form: 
\begin{equation}
{\rm real}=\alpha _{S}\int_{0}^{E_{X}}\int_{0}^{1}\frac{d\epsilon }{\epsilon 
}\frac{d\theta ^{2}}{\theta ^{2}}\, \, \delta \left( \epsilon \theta ^{2}-%
\frac{m_{X}^{2}}{E_{X}}\right)  \label{sudar}
\end{equation}
and 
\begin{equation}
{\rm virtual}=-\alpha _{S}\,\delta \left( \frac{m_{X}^{2}}{E_{X}}\right)
\int_{0}^{E_{X}}\int_{0}^{1}\frac{d\epsilon }{\epsilon }\frac{d\theta ^{2}}{%
\theta ^{2}},  \label{sudav}
\end{equation}
where $\epsilon$ is the gluon energy, $\theta $ is the angle between the $up$
and the gluon, and $\Theta=\pi - \theta$ is the polar angle of the gluon
3-momentum. The perturbative corrections of the form (\ref{largelog}) blow
up at the Born kinematics $m_{X}=0$, which is the threshold of the inelastic
channels. For this reason, the above corrections are often called threshold
logarithms and need a resummation to any order in $\alpha _{S}$.

\section{ Overview of Factorization}

\label{strategy}

The aim of this paper is a detailed study of factorization in semi-inclusive
heavy flavour decays and of the properties of the shape function in field
theory. In order to trace all the perturbative and non-perturbative
contributions to the process, it is convenient to perform the factorization
in two steps. In the first step the heavy flavour is replaced by a static
quark. That is accomplished by taking the infinite mass limit (\ref{hqet}),
keeping $E_{X}$ and $m_{X}$ fixed. With this, the hadronic tensor loses a
kinematical scale, namely the heavy flavour mass $m_{Q}$: 
\begin{equation}
W_{\mu \nu }\left( m_{Q},E_{X},m_{X}\right) \rightarrow \widetilde{W}_{\mu
\nu }\left( E_{X},m_{X}\right) ,  \label{twotens}
\end{equation}
where the effective hadronic tensor is defined as 
\begin{equation}
\widetilde{W}_{\mu \nu }\ \equiv \sum_{X}\,\langle H_{Q}|\widetilde{J}_{\nu
}^{\dagger }|X\rangle \langle X|\widetilde{J}_{\mu }|H_{Q}\rangle \,\delta
^{4}(p_{B}-q-p_{X}),  \label{Wstep1}
\end{equation}
and it contains the static-to-light currents 
\begin{equation}
\widetilde{J}_{\mu }(x)=\overline{q}(x)\Gamma _{\mu }\widetilde{Q}(x).
\end{equation}
The difference between the two tensors in eq.~(\ref{twotens}) is
incorporated into a first coefficient function or hard factor. While in full
QCD the vector and axial currents are conserved, or partially conserved, so
the renormalization constants are UV-finite and anomalous dimensions vanish,
this property does not hold anymore in the HQET: the effective current with
a static quark is not conserved and it acquires an anomalous dimension $%
\widetilde{\gamma _{J}}$~\footnote{%
~In eq.~(\ref{simbolo_1}) and (\ref{simbolo_2}), we are representing the
evolution schematically, without details; f.i., we do not distinguish
between the anomalous dimensions of the vector and axial currents.}\cite
{polwise}: 
\begin{equation}
\left( \frac{d}{d\log \mu }+\widetilde{\gamma _{J}}\right) \langle 
\widetilde{J}_{\mu }\rangle =0.  \label{simbolo_1}
\end{equation}
As a consequence also the hadronic tensor acquires an anomalous dimension,
which equals twice that of the vector or axial current: 
\begin{equation}
\left( \frac{d}{d\log \mu }+2\widetilde{\gamma _{J}}\right) \widetilde{W}%
_{\mu \nu }=0.  \label{simbolo_2}
\end{equation}
All this is very easily understood by observing that the original QCD tensor 
$W_{\mu \nu }$ is UV-finite at one loop but it does contain $\alpha _{S}\log
\,m_{Q}$ terms, and so it is divergent in the infinite-mass limit (\ref{hqet}%
). If this limit is taken {\it ab initio, }i.e. before regularization, these
terms manifest themselves as new ultraviolet divergences, an heritage of the 
$\log \,m_{Q}$ terms of the original tensor. We may say that the dependence
on the heavy mass is promoted to UV divergence; in practice 
\begin{equation}
\alpha _{S}\,\log \,\frac{m_{Q}}{E_{X}}\rightarrow \alpha _{S}\,\log \frac{%
\Lambda _{1}}{E_{X}},
\end{equation}
where $\Lambda _{1}$ is an UV cutoff \ if we deal with the bare theory, or a
renormalization point if we deal with the renormalized theory; in principle $%
\Lambda _{1}\ll m_{Q}$. At the end of the game, the effective hadronic
tensor still depends on three scales, just like the original one, 
\begin{equation}
\widetilde{W}_{\mu \nu }=\widetilde{W}_{\mu \nu }\left( E_{X},m_{X};\Lambda
_{1}\right) .
\end{equation}
The original tensor $W_{\mu \nu }$ is parametrized in terms of five
independent form factors \cite{neubfaz}. For the HQET hadronic tensor (\ref
{Wstep1}) there are instead relations between the form factors originating
from the spin-symmetry of the HQET. In particular, the structure in $\log
m_{Q}/E_{X}$ of the original QCD tensor can be understood by looking at the
UV divergences of \ $\widetilde{W}_{\mu \nu }~\footnote{%
In Dimensional Regularization (DR), this means simple poles $1/\epsilon $.}$.

After the first step $\widetilde{W}_{\mu \nu }$ still contains perturbative
contributions. The latter are factorized with a second step, which
corresponds to the limit (\ref{one}). Additional UV divergences are
introduced also with this second step, which must be regulated with a new
cutoff $\Lambda _{2}$. In principle $\Lambda _{2}\ll E_{X}$, since $%
E_{X}\rightarrow \infty $. As before with the heavy mass logarithms, soft
and collinear logarithms are promoted to ultraviolet logarithms: 
\begin{equation}
\alpha _{S}\left( \frac{\log m_{X}^{2}/E_{X}^{2}}{m_{X}^{2}/E_{X}^{2}}%
\right) _{+}\rightarrow \alpha _{S}\left( \frac{\log \left( -2k_{+}/\Lambda
_{2}\right) }{-2k_{+}/\Lambda _{2}}\right) _{+}.
\end{equation}
The second factorization step involves double-logarithmic effects of an
infrared nature, in contrast with the single logarithms of the large mass of
the first step. In practice, we separate perturbative contributions from
non-perturbative ones starting with a cutoff 
\begin{equation}
\Lambda _{2}\sim E_{X}  \label{uppe}
\end{equation}
and lowering it to a much smaller value\footnote{%
In order to avoid substantial finite cutoff effects, the condition $\Lambda
_{2}^{\prime }\gg \Lambda _{QCD}$ must hold.} 
\begin{equation}
\Lambda _{2}^{\prime }\ll E_{X}.  \label{lowe}
\end{equation}
The contributions of the fluctuations with energy between $\Lambda _{2}$ and 
$\Lambda _{2}^{\prime }$ are put into a second coefficient function, while
the contributions below $\Lambda _{2}^{\prime }$ are factorized inside the
shape function. The latter is defined in the framework of a low-energy
effective theory, with a cutoff given by 
\begin{equation}
\Lambda _{ET}=\Lambda _{2}^{\prime }.  \label{basso}
\end{equation}
Most of the non-perturbative effects in lattice-like regularizations are
contained in the shape function, which uniquely determines the final,
non-perturbative, hadronic tensor 
\begin{equation}
\widetilde{\widetilde{W}}_{\mu \nu }\ \equiv \sum_{X}\,\langle H_{Q}|%
\widetilde{\widetilde{J}}_{\nu }^{\dagger }|X\rangle \langle X|\widetilde{%
\widetilde{J}}_{\mu }|H_{Q}\rangle \,\delta ^{4}(p_{B}-q-p_{X}),
\label{tiltil}
\end{equation}
containing the effective-heavy-to-effective-light currents 
\begin{equation}
\widetilde{\widetilde{J}}_{\mu }(x)=\overline{\widetilde{q}}(x)\Gamma _{\mu }%
\widetilde{Q}(x).
\end{equation}
It is worth noting that the tensor (\ref{tiltil}) involves a single form
factor, proportional to the shape function itself (see eq. (\ref{singleff}%
)). That is again a consequence of the spin-symmetry of both HQET and 
LEET~\cite{francesi}, which is more efficient than
that one of the HQET alone. The shape function is completely
non-perturbative and perturbative factors can no longer be extracted.

The effect of lowering the UV cutoff (eqs.~(\ref{uppe}) and (\ref{lowe})) is
incorporated inside a coefficient function, which, unlike more simple cases
such as the light-cone expansion in DIS, is not completely short-distance
dominated. Some long-distance effects are left in the coefficient function,
but they are expected to be suppressed on physical grounds. Finally, the
introduction of ultraviolet divergences with factorization, implies
scheme-dependence issues for the shape function, which are rather dramatic
because of the double-logarithmic nature of the problem (cf. eqs.(\ref{sudar}%
) and (\ref{sudav})).

In fig. 1, we give a pictorial description of the above procedure.

\section{OPE}

\label{opesection}

The amplitude for the decay (\ref{semilep}), which we take as our example
from now on, can be written at the lowest order in the weak coupling as 
\begin{equation}
A=\frac{G_{F}}{\sqrt{2}}\,\langle l\nu \mid L_{\mu }\mid 0\rangle \,\langle
X|J^{\mu }|B\rangle,  \label{ampli}
\end{equation}
where $L_{\mu }$ is the leptonic current and $J_{\mu }$ is the hadronic one: 
\begin{equation}
J_{\mu }(x)\equiv \overline{q}(x) \Gamma_\mu b(x)
\end{equation}
with $\Gamma_\mu = \gamma_{\mu }(1-\gamma_{5})$, $q(x)$ a light quark field
and $b(x)$ the beauty quark field. Taking the square of (\ref{ampli}) and
summing over the final states, we arrive at the hadronic tensor defined in
eq.~(\ref{Wmunu}). By the optical theorem, we can relate the hadronic tensor 
$W_{\mu \nu }$ to the imaginary part\footnote{%
Since $\Gamma_\mu$ is in general complex, we should say, more properly, the
absorptive part.} of the Green function or forward hadronic tensor $T_{\mu
\nu }$: 
\begin{equation}
W_{\mu \nu }=-\frac{1}{\pi }{\rm Im}\,\,T_{\mu \nu },  \label{optical}
\end{equation}
where 
\begin{equation}
T_{\mu \nu }\ \equiv -i\int d^{4}x\,e^{-iqx}\langle B|T\left( J_{\mu
}^\dagger(x)J_{\nu }(0)\right) |B\rangle .  \label{tensorT}
\end{equation}

\subsection{The HQET}

\label{opesection_1}

We are interested in the evaluation of $T_{\mu \nu }$ in the effective
theory and we discuss in this section the first factorization step:
replacing the beauty quark by a quark in the HQET. As is well known, we can
decompose the heavy quark field $b(x)$ into two effective quark and
antiquark fields $h_{v}$ and $H_{v}$~\footnote{%
We prefer to refer to the physical $B$-meson mass rather than to the
unphysical $b$-quark mass. Their difference is of order $\Lambda _{QCD}$, so
it is a $1/m_{B}$ correction and can be neglected in our leading-order
analysis. Furthermore, in perturbation theory there is no binding energy so
that $m_{b}=m_{B}$.} 
\begin{equation}
b(x)=e^{-im_{B}v\cdot x}\left[ h_{v}(x)+H_{v}(x)\right] ,
\end{equation}
satisfying 
\begin{equation}
P_{+}\ h_{v}=h_{v},\ \ P_{-}\ h_{v}=0,\ \ \ P_{-}\ H_{v}=H_{v},\ \ P_{+}\
H_{v}=0,
\end{equation}
where $P_{\pm }=(1\pm \hat{v})/2$ are the projectors over the components
with positive and negative energies, respectively. The field $H_{v}$ is
neglected (which amounts to neglecting heavy-pair creation), so that 
\begin{equation}
b(x)\simeq e^{-im_{B}v\cdot x}\ h_{v}(x).  \label{quarkH}
\end{equation}
By using eq.~(\ref{quarkH}) we obtain 
\begin{equation}
\widetilde{T}_{\mu \nu }=-i\int d^{4}x\ e^{iQ\cdot x}\langle B(v)|T\ \bar{h}%
_{v}(x)\Gamma _{\mu }^{\dagger }\ q(x)\ \bar{q}(0)\Gamma _{\nu }\
h_{v}(0)|B(v)\rangle .
\end{equation}
We now use the Wick theorem and we single out the only contraction that is
relevant to $B$ decay: 
\begin{equation}
\widetilde{T}_{\mu \nu }=\int d^{4}x\ e^{iQx}\langle B|\bar{h}_{v}(x)\Gamma
_{\mu }^{\dagger }\,S(x|0)\Gamma _{\nu }h_{v}(0)|B\rangle ,  \label{interm}
\end{equation}
where $S(x|0)$ is the light quark propagator. Note that the operator
entering the right-hand side of eq.~(\ref{interm}) is already normal
ordered, since $h_{v}$ has only the component that annihilates heavy quarks,
while $\bar{h}_{v}$ only the components that create them. We can express the
Fourier transform of the light quark propagator as\footnote{%
The notation is very compact. For more explicit representations of the
propagator see ref.~\cite{concorbo}.} 
\begin{equation}
S(Q+iD)=\frac{1}{i\hat{D}+\hat{Q}+i0}=\frac{i\hat{D}+\hat{Q}}{Q^{2}+2iD\cdot
Q-D^{2}-g/2\ \sigma _{\mu \nu }G^{\mu \nu }+i0}  \label{propagatore_1}
\end{equation}
where $\sigma _{\mu \nu }\equiv i/2[\gamma _{\mu },\gamma _{\nu }]$ is a
generator of the Lorentz group, $G_{\mu \nu }\equiv -i/g[D_{\mu },D_{\nu }]$
is the field strength and $D_{\mu }\equiv \partial _{\mu }-igA_{\mu }$ is
the covariant derivative. In eq. (\ref{propagatore_1}), $0$ denotes, as
usual, an infinitesimal positive number and gives the prescription to deal
with pole or branch-cut singularities. There are three different regions
according to the value of the jet invariant mass, which are described by
three different full or effective theories: 
\begin{eqnarray}
i)\qquad m_{X}^{2} &\sim &O(E_{X}^{2}),  \nonumber \\
ii)\qquad m_{X}^{2} &\sim &O(\Lambda _{QCD}^{2}\,),  \nonumber \\
iii)\qquad m_{X}^{2} &\sim &O(\Lambda _{QCD}\,E_{X}).
\end{eqnarray}
Since the derivative of the rescaled $h_{v}$ field brings down the residual
momentum $k^{\prime }$, and it is therefore an operator with matrix elements
of order $O\left( \Lambda _{QCD}\right) $, the matrix elements of the
operators entering the light quark propagator have a size of the order of 
\begin{eqnarray}
\langle \,i\hat{D}\,\rangle &\sim &O(\Lambda _{QCD}),  \nonumber \\
\langle \,2iD\cdot Q\ \rangle &\sim &O(\Lambda _{QCD}\,E_{X}),  \nonumber \\
\langle \,D^{2}\,\rangle &\sim &O(\Lambda _{QCD}^{2}),  \nonumber \\
\langle \,\sigma _{\mu \nu }G^{\mu \nu }\,\rangle &\sim &O(\Lambda
_{QCD}^{2}).  \label{estim}
\end{eqnarray}

Let us discuss these regions in turn in the next section.

\subsection{ General kinematical regions}

\begin{description}
\item[$i)$]  This region corresponds to a jet $X$ with a large invariant
mass, of the order of the energy: 
\begin{equation}
m_{X}\sim O\left( E_{X}\right) \,.
\end{equation}
To a first approximation all the covariant derivative terms can be
neglected, so that 
\begin{equation}
S(Q+iD)\ \simeq \ \frac{\hat{Q}}{Q^{2}+i0},
\end{equation}
i.e. the light quark can be described as a free quark. A higher accuracy is
reached when expanding the propagator in powers of the covariant derivative
operators up to the required order. We have here an application of the $%
1/m_{B}$ expansion up to a prescribed (finite) order \footnote{%
It is clear that a consistent inclusion of the $1/m_{B}$ corrections
involves also the expansion of the heavy quark field $b(x)$ into the
effective quark field $h_{v}(x)$ up to the required order.}. In this region
there are no large adimensional ratios of scales, the latter being all of
the same order. This implies that in perturbation theory we do not hit large
logarithmic corrections to be resummed to all orders in $\alpha _{S}.$ This
region is not relevant to the endpoint electron spectrum because the
hadronic jets takes away most of the available energy. This region will not
be discussed further here.

\item[$ii)$]  This region involves a recoiling hadronic system with a mass
of the order of the hadronic one: it can be a single hadron or very few
hadrons. The dynamics is dominated by the emission, with consequent decay,
of few resonances; it is a completely non-perturbative problem. According to
the estimates (\ref{estim}), no term can be neglected in the light quark
propagator. We are faced with full QCD dynamics as far as the final hadronic
state is concerned. This region must be evaluated by an explicit sum over
all the kinematically possible hadronic states, and the latter have to be
computed with a non-perturbative technique such as a quark model or lattice
QCD. This region will not be discussed here either.

\item[$iii)$]  This region is intermediate between $i)$ and $ii)$ and as
such it has both perturbative and non-perturbative components. Roughly
speaking, we have to take into account non-perturbative effects for the
initial state hadron, while we can neglect final state binding effects. This
region is characterized by a small ratio of the jet invariant mass to the
jet energy, and thus involves the large adimensional ratio in (\ref{smallra}%
). As always is the case, perturbation theory generates logarithms of the
above adimensional ratio, eq.~(\ref{largelog}). The term $2\, i D\cdot Q$ at
the denominator cannot be brought at the numerator (with a truncated
operatorial expansion) because it is of the same order as $m_{X}^{2}.$ At
lowest order, the other covariant derivative terms can be neglected, to
give: 
\begin{equation}
S(Q+iD)\simeq \frac{\hat{Q}}{m_{X}^{2}+2iD\cdot Q+i0 }.  \label{leetpro}
\end{equation}
One can reach a higher level of accurary keeping these latter corrections up
to a given order\footnote{%
We envisage a relation between the $1/E_{X}$ corrections to the shape
function and the power-suppressed perturbative corrections of the form $%
\alpha _{S}/E_{X}\, \log ^{2}(E_{X}/m_{X})$.}. The rest of the paper deals
with region $iii)$ at the lowest order in $1/m_{B}$.
\end{description}

\subsection{The LEET}

In this section we discuss the second factorization step, which involves the
description of the final $up$ quark in the LEET, according to eq.~(\ref
{leetpro}). Let us define the adimensional vector $n_{\mu }$ as: 
\begin{equation}
n_{\mu }=\frac{Q_{\mu }}{Q\cdot v}~.
\end{equation}
This $n_{\mu }$ has a normalized time component, $n_{0}=1$. In the
``semi-inclusive'' endpoint region $iii)$: 
\begin{eqnarray}
n^{2}~ &=&\frac{m_{X}^{2}}{E_{X}^{2}}  \nonumber \\
&=&O\left( \frac{\Lambda _{QCD}}{E_{X}}\right) ~\ll 1.  \label{nquadro}
\end{eqnarray}
We will show later that $n$ can be replaced by a vector lying exactly on the
light-cone, i.e. 
\begin{equation}
n\rightarrow \overline{n},
\end{equation}
where $\overline{n}^{\mu }=(1;0,0,-1)$ (\thinspace $\overline{n}^{\,2}=0$),
representing the direction of the hadronic jet, the $-z$ axis. We can write 
\begin{equation}
S(Q+iD)=\frac{1}{2v\cdot Q}\;\ \frac{\hat{Q}}{iD_{+}-k_{+}+i0},
\end{equation}
where $k_{+}$ has been defined in eq.~(\ref{defkplus}) and $D_{+}\equiv 
\overline{n}\cdot D.$ We can simplify the tensor structure of $T_{\mu \nu }$
by using the identity 
\begin{equation}
\bar{h}_{v}\Gamma _{\mu }h_{v}=\frac{1}{2}{\rm Tr}(\Gamma _{\mu }P_{+})\ 
\bar{h}_{v}h_{v}-\frac{1}{2}{\rm Tr}(\gamma _{\mu }\gamma _{5}P_{+}\Gamma
_{\mu }P_{+})\,\bar{h}_{v}\gamma ^{\mu }\gamma _{5}h_{v},
\end{equation}
which is valid for any $\Gamma _{\mu }$. The matrix element of the axial
vector current between the $B$-meson states vanishes by parity invariance,
so that \footnote{%
A physical argument for the spin factorization is that, in the limit $%
m_{B}\rightarrow \infty $, the spin interaction of the $b$-quark in the $B$
meson vanishes; therefore we can average over the helicity states of the $b$
quark \cite{memasse}.}: 
\begin{equation}
\widetilde{\widetilde{T}}_{\mu \nu }=\;s_{\mu \nu }\;\frac{1}{2v\cdot Q}\
F(k_{+}),
\end{equation}
where we have defined the ``light-cone'' function 
\begin{equation}
F(k_{+})\equiv \langle B(v)\mid h_{v}^{\dagger }\frac{1}{iD_{+}-k_{+}+i0}%
h_{v}\mid B(v)\rangle ,  \label{defFhqet}
\end{equation}
and 
\begin{equation}
s_{\mu \nu }\equiv \frac{1}{2}\,{\rm Tr}\left[ \Gamma _{\mu }^{\dagger }\hat{%
Q}\Gamma _{\nu }P_{+}\right] \;
\end{equation}
is the ``spin factor'', containg the leading spin effects. The factor $%
1/(2v\cdot Q)$ is a jacobian, which appears as we go from the full QCD
variable $Q^{2}$ to the effective theory variable $k_{+}.$

Taking the imaginary part of $T_{\mu \nu }$, we obtain (see relation (\ref
{optical})): 
\begin{equation}
\widetilde{\widetilde{W}}_{\mu \nu }=\;s_{\mu \nu }\;\frac{1}{2v\cdot Q}\
f(k_{+}),  \label{singleff}
\end{equation}
where 
\begin{equation}
f(k_{+})\equiv -\frac{1}{\pi }{\rm Im}\,F(k_{+})
\end{equation}
is the shape function. By using the formula 
\begin{equation}
\frac{1}{iD_{+}-k_{+}+i0}={\rm P}\frac{1}{i\,D_{+}-k_{+}}-i\pi \,\delta
(iD_{+}-k_{+}),
\end{equation}
we recover the definition of the shape function given by eq.~(\ref{defshape}%
). Note that it involves the non-local operator $h_{v}^{\dagger }\,\delta
(k_{+}-iD_{+})\,h_{v}$, which results from the resummation of the towers of
operators of the form $\left( Q\cdot D\right) ^{n}$.

\subsection{The Variable Mass}

\label{variable_mass}

The hadronic tensor can be written in the effective theory in terms of the
shape function as: 
\begin{equation}
\widetilde{\widetilde{W}}_{\mu \nu }=\ s_{\mu \nu }\int_{-\infty
}^{0}dk_{+}\ \delta \left( Q^{2}+2k_{+}v\cdot Q\right) \,f(k_{+});
\end{equation}
in the second member, $k_{+}$ is an integration, i.e. dummy, variable. In
the free theory, with an on-shell $b$-quark (i.e. $k^{\prime }=0$ in eq. (%
\ref{momhq})), 
\begin{equation}
f^{0}(k_{+})=\delta (k_{+}-0),
\end{equation}
so that 
\begin{equation}
W_{\mu \nu }^{0}=s_{\mu \nu }\ \delta (Q^{2}-0).
\end{equation}
The hadronic tensor can be written, up to terms of order $k_{+}^{2}\sim
O\left( \Lambda _{QCD}^{2}\right) ,$ as 
\begin{equation}
\widetilde{\widetilde{W}}_{\mu \nu }=\ s_{\mu \nu }(Q)\int dm_{\ast }\
\delta \left( Q_{\ast }^{2}-0\right) \,f(m_{\ast }-m_{B}),
\end{equation}
where we have defined 
\begin{equation}
Q_{\ast }\equiv m_{\ast }v-q
\end{equation}
and $m_{\ast }$ is the ``variable'' or ``fragmentation'' mass, defined in
eq.(\ref{massvar}). Since $m_{\ast }$ is just a shift of $k_{+}$, the range
is 
\begin{equation}
-\infty <m_{\ast }\leq m_{B}.
\end{equation}
Inside $s_{\mu \nu }$ we can replace $Q$ with $Q_{\ast },$ because that
amounts only to a correction of order $k_{+}=O(\Lambda _{QCD})$, so that%
\footnote{%
We replace by 0 the lower limit of integration, because the relevant region
is $m_{\ast }\sim m_{B}-O(\Lambda _{QCD})$.} 
\begin{equation}
\widetilde{\widetilde{W}}_{\mu \nu }(v,Q)=\ \int_{0}^{m_{B}}dm_{\ast }\
\varphi (m_{\ast })\,\ W_{\mu \nu }^{0}(v,Q_{\ast }),  \label{funda}
\end{equation}
where 
\begin{equation}
W_{\mu \nu }^{0}(v,Q^{\ast })=\ s_{\mu \nu }(Q_{\ast })\ \delta \left(
Q_{\ast }^{2}-0\right)
\end{equation}
is the hadronic tensor in the free theory for a heavy quark of mass $m_{\ast
}$ and 
\begin{equation}
\varphi (m_{\ast })\equiv f(m_{\ast }-m_{B})
\end{equation}
is the distribution for the effective mass $m_{\ast }$ of the $b$-quark
inside the $B$-meson at disintegration time. Equation~(\ref{funda}) is the
fundamental result of factorization in semi-inclusive heavy flavour decays:
it says that the hadronic tensor in the effective theory can be expressed as
the convolution of the hadronic tensor in the free theory with a variable
mass times a distribution probability for this mass. That offers also the
physical interpretation to the shape function anticipated in the
introduction: it represents the probability that the $b$ quark has an
effective mass $m_{\ast }$ at the decay time. Since this tensor encodes all
the hadron dynamics, $any$ distribution can be expressed in a similar
factorized form.

\section{Factorization in the quantum theory}

\label{factorization_quantum}

In this section we discuss factorization in the quantum theory, i.e. the
separation of short-distance and long-distance contributions, including loop
effects.

A shape function $f(k_{+})^{QCD}$ and a light-cone function $F(k_{+})^{QCD}$
can also be defined in full QCD by means of the relations~\cite{noi}: 
\[
T_{\mu \nu }^{QCD}\equiv \left( s_{\mu \nu }\;+\Delta s_{\mu \nu }\right) 
\frac{1}{2v\cdot Q}\ F(k_{+})^{QCD} 
\]
and 
\[
W_{\mu \nu }^{QCD}\equiv \left( s_{\mu \nu }\;+\Delta s_{\mu \nu }^{\prime
}\right) \frac{1}{2v\cdot Q}\ f(k_{+})^{QCD}, 
\]
where $\Delta s_{\mu \nu }$ and $\Delta s_{\mu \nu }^{\prime }$ are defined
as the part of the spin structure not proportional to $s_{\mu \nu }$. The
tensors $\Delta s_{\mu \nu }$ and $\Delta s_{\mu \nu }^{\prime }$ represent
residual spin effects not described by the effective theory (ET), which do
not contribute to the Double-Logarithmic Approximation (DLA)\footnote{%
Note that $\Delta s_{\mu \nu }$ and $\Delta s_{\mu \nu }^{\prime }$ are, in
general, different; this was not noted in~\cite{noi}.}. In DLA the forward
tensor can therefore be written as 
\begin{equation}
T_{\mu \nu }^{QCD}=\;s_{\mu \nu }\;\frac{1}{2v\cdot Q}\ F(k_{+})^{QCD},
\label{Tforward}
\end{equation}
where the ``light-cone function'' is given by 
\begin{equation}
F(k_{+})^{QCD}\equiv \frac{1}{-k_{+}+i0}\left[ 1+a\,C\right] ,
\label{Fgrande}
\end{equation}
$a\equiv \alpha _{S}$\bigskip $C_{F}/\pi $ and $C$ is the scalar triangle
diagram (see fig. 2): 
\begin{equation}
C\equiv \ -i\,v\cdot Q\int \frac{d^{4}l}{\pi ^{2}}\ \frac{1}{(l+Q)^{2}+i0}~\ 
\frac{1}{v\cdot l+l^{2}/2m+i0}~\ \frac{1}{l^{2}+i0}.
\end{equation}
We have set the light quark mass equal to zero \cite{giulia}.

The hadronic tensor relevant to the decay is obtained by taking the
imaginary part according to eq.~(\ref{optical}). This transforms the
products in convolutions, which are converted again into ordinary products
by the well-known Mellin transform \cite{cattren}.

Infrared singularities (soft \& collinear) are regulated by the virtuality $%
Q^{2}\neq 0$ of the external $up$ quark\footnote{%
This is consistent because a virtual massless quark is not degenerate with a
quark and a soft and/or collinear gluon.}. We may write 
\begin{equation}
Q^{\mu }\cong E_{X}\left( 1+\frac{n^{2}}{4};0,0,-1+\frac{n^{2}}{4}\right)
=E_{X}\,\left( v_{-}+\frac{n^{2}}{4}v_{+}\right)  \label{defQbig}
\end{equation}
and 
\begin{equation}
n^{\mu }\cong v_{-}^{\mu }+\frac{n^{2}}{4}v_{+}^{\mu }\,,\qquad \overline{n}%
^{\,\mu }=v_{-}^{\mu }
\end{equation}
where we defined the light-cone versors 
\begin{equation}
v_{+}\equiv \left( 1;0,0,1\right) ,\qquad v_{-}\equiv \left( 1;0,0,-1\right)
.
\end{equation}
Let us now consider the properties of the integral $C$. First, it is
adimensional. Second, it is UV-finite for power counting: the integrand has
three ordinary scalar propagators with a total of six powers of momentum at
the denominator. This implies that $C$ does not depend on an ultraviolet
cutoff $\Lambda _{UV}$ as long as it is larger than any physical scale of
the process, namely 
\begin{equation}
\Lambda _{UV}\gg m_{B}.  \label{largecut}
\end{equation}
Third, as already discussed, $C$ is also IR-finite as long as $Q^{2}\neq 0.$
For $Q^{2}>0$ there is an imaginary part, related to the propagation of the
real $up$ and gluon pair, while for $Q^{2}<0$ the integral is real.
Therefore $C$ does depend on adimensional ratios of three different scales: $%
m_{B},E_{X}$ and $m_{X}.$ There are only two independent ratios, which we
choose as $m_{B}/E_{X}$ and$\,\,m_{X}/E_{X}.$ We are going to decompose the
integral $C$ in a sum of various integrals; at the end, one of them will
correspond to the double-logarithmic contribution to the shape function $%
f(k_{+})$ in the low-energy ET. The other integrals represent additional
contributions and they are mostly short-distance dominated in lattice-like
regularizations. This decomposition consists of two separate steps, which
will be described in the following sections.

\subsection{From QCD to HQET}

In the first step we isolate a hard factor by simply subtracting and adding
back the integral with the full beauty quark propagator replaced by a static
one (see fig. 3) 
\begin{equation}
C=C_{s}+C_{h},
\end{equation}
where 
\begin{equation}
C_{s}\equiv -iv\cdot Q\int \frac{d^{4}l}{\pi ^{2}}\ \frac{1}{(l+Q)^{2}+i0}~\ 
\frac{1}{v\cdot l+i0}~\ \frac{1}{l^{2}+i0},  \label{defC0}
\end{equation}
and 
\begin{equation}
C_{h}\equiv i\frac{v\cdot Q}{2m}\int \frac{d^{4}l}{\pi ^{2}}\ \frac{1}{%
(l+Q)^{2}+i0}\ \frac{1}{v\cdot l+i0}\,\frac{1}{v\cdot l+l^{2}/2m+i0}.
\end{equation}
The above decomposition parallels that one performed in section~\ref
{opesection} in operatorial language. The light-cone function factorizes at
order $\alpha _{S}$ according to 
\begin{equation}
F(k_{+})^{QCD}=\frac{1}{-k_{+}+i0}\left[ 1+a\,C\right] =\frac{1}{-k_{+}+i0}%
\left[ 1+a\,C_{h}\right] \left[ 1+a\,C_{s}\right] .  \label{LCfunction}
\end{equation}
We expect that $C_{s}$ has the same infrared behaviour as $C$; it will be
subjected to a further decomposition in the next sections; $C_{h}$ \ is the
``hard factor'', i.e. the difference between QCD and the static limit for
the $b$ quark. The latter integral is both UV- and IR-finite. The
ultraviolet finiteness stems from power counting: the integrand has two
scalar propagators and a static propagator with a total of five powers of
momentum in the denominator~\footnote{%
It is known that ultraviolet power counting may fail in effective theory
integrals when there are LEET propagators because of the occurrence of an
ultraviolet collinear region~\cite{noi}. The integral $C_{h}$, however,
contains only an HQET propagator.}. In the infrared region, all the
components of the loop momentum are small 
\begin{equation}
{\rm IR}:\quad l_{\mu }\rightarrow \rho l_{\mu },\quad \rho \rightarrow 0,
\end{equation}
so we can neglect the terms that are quadratic in $l^{\mu }$ in the
propagator denominators: 
\begin{equation}
C_{h\,,IR}\sim \int d^{4}l\frac{1}{2l\cdot Q+Q^{2}+i0}~~\ \frac{1}{(v\cdot
l+i0)^{2}}.
\end{equation}
Integrating over $l_{0}$, and closing the contour in the upper half of the $%
l_{0}$-plane, we see that there are no enclosed poles and the integral
vanishes (QED). Inside $C_{h}$ we can therefore make the replacement~%
\footnote{%
With this substitution, terms related to higher twist contributions of the
forms $\left( m_{X}/m_{B}\right) ^{n_{1}}$ and $\left( m_{X}/E_{X}\right)
^{n_{2}}$ are neglected, but this is in agreement with our leading-twist
ideology (the indices $n_{1}$ and $n_{2}$ are integers).} 
\begin{equation}
Q^{\mu }\quad \rightarrow \quad \overline{Q}^{\,\mu }\equiv
E_{X}\,v_{-}\qquad \qquad (\overline{Q}^{\,2}=0).
\end{equation}
It follows that $C_{h}$ depends only on $m_{B}$ and $E_{X}$: $%
C_{h}=C_{h}(m_{B},E_{X}).$ Since it is adimensional, it may depend only on
the adimensional hadronic energy 
\begin{equation}
z\equiv \frac{2E_{X}}{m_{B}},
\end{equation}
i.e. $C_{h}=C_{h}(z).$ An explicit computation gives 
\begin{equation}
C_{h}=\log \left( 1-z\right) \log z+Li_{2}(z)\simeq -z\log z\quad \left(
z\ll 1\right) ,
\end{equation}
$C_{h}$ does not contain large logarithmic contributions in the limit $%
z\rightarrow 0$, i.e. $\log z$ terms \footnote{%
The dilogarithm is defined as
\par
\begin{equation}
Li_{2}(z)\equiv -\int_{0}^{z}\frac{\log (1-x)}{x}dx=\sum_{n=1}^{\infty }%
\frac{z^{n}}{n^{2}}\quad \left( |z|\leq 1\right) .
\end{equation}
}. This is related to the fact that $C_{h}$ and $C_{s}$ are UV-convergent.
In general, single logarithms of the hadronic energy, $\log z,$ do appear,
representing the difference between the interaction of a full propagating
heavy quark, of mass $m_{B},$ and that one of a static quark. The relevant
interaction energies are between the beauty mass $m_{B}$ and the hadronic
energy $E_{X}$, 
\begin{equation}
\alpha _{S}\,\int_{z^{2}m_{B}^{2}}^{m_{B}^{2}}\frac{dk^{2}}{k^{2}}%
=-2\,\alpha _{S}\,\log z.  \label{logz}
\end{equation}
The logarithms (\ref{logz}) are resummed, as usual, by replacing the bare
coupling with the running coupling and exponentiating, so that the above
formula is corrected into: 
\begin{eqnarray}
1+\gamma _{0}\alpha _{S}\int_{z^{2}m_{B}^{2}}^{m_{B}^{2}}\frac{dk^{2}}{k^{2}}%
\quad &\rightarrow &\quad \exp \left[ \gamma
_{0}\int_{z^{2}m_{B}^{2}}^{m_{B}^{2}}\frac{dk^{2}}{k^{2}}\,\alpha _{S}(k^{2})%
\right]  \nonumber \\
&=&\exp \left[ -2\,\alpha _{S}\gamma _{0}\log z+2\gamma _{0}\beta
_{0}\,\alpha _{S}^{2}\,\log ^{2}z+\cdots \right] ,\quad
\end{eqnarray}
where $\beta _{0}\equiv 1/(4\pi )(11/3\,N_{c}-2/3\,n_{f})$ and $\alpha
_{S}\equiv \alpha _{S}(m_{B})$.

Let us summarize the above discussion. A first coefficient function is
introduced, which takes into account the fluctuations with energy $%
\varepsilon $ in the range 
\begin{equation}
E_{X}<\varepsilon <m_{B}.
\end{equation}
In the language of Wilson's renormalization group, we are lowering the UV
cutoff of the effective hamiltonian from $m_{B}$ to $E_{X}$.

\subsection{ Infrared factorization}

The second factorization step involves the separation of the various
infrared contributions to the process, one of which will ultimately lead to
the shape function. This step forces us to introduce explicitly an
ultraviolet cutoff $\Lambda $ from which the various factors depend
separately. In other words, the decomposition of $C_{s}$ introduces
fictitious UV divergences, which cancel in the sum. As anticipated in the
overview section, infrared factorization in our scheme involves two
different operations:

\begin{itemize}
\item  the separation of the various pole contributions to the QCD amplitude
according to the Cauchy theorem;

\item  the lowering of the UV cutoff from $\Lambda _{UV}\gtrsim E_{X}$ to $%
\Lambda _{UV}=\Lambda _{ET}\ll E_{X},$where $\Lambda _{ET}$ is the cutoff of
the final low-energy effective theory, in which the shape function is
defined (the latter contains the majority of the non-perturbative effects).
\end{itemize}

Step 1) will be discussed in this section, while step 2) is treated mostly
in the next section. 

$C_{s}$ is UV-convergent, as is clear again from power counting, and it does
not depend on the beauty mass $m_{B}$, which has been sent to infinity, so
that $C_{s}=C_{s}(E_{X},m_{X}).$ The only adimensional variable that can be
constructed out of $E_{X}$ and $m_{X}$ is their ratio or, equivalently, $%
n^{2}$ (see eq.~(\ref{nquadro})). Since $C_{s}$ is adimensional it may
depend only on $n^{2}$: $C_{s}=C_{s}(n^{2}).$ The explicit computation in
DLA gives 
\begin{equation}
C_{s}=-\frac{1}{2}\log ^{2}(-n^{2}-i0)\qquad \quad \quad \quad \left(
DLA\right) .  \label{calcolo_cs}
\end{equation}
The infrared factorization is performed by integrating $C_{s}$ over the
energy $l_{0}$ using the Cauchy theorem. There are three poles in the lower
half of the $l_{0}$-plane related to the propagation of a {\it real} static
quark, a {\it real }gluon and a {\it real }$up$ quark, located respectively
at 
\begin{equation}
l_{0}=-i0,\qquad l_{0}=+|\overrightarrow{l}|-i0,\qquad l_{0}=-Q_{0}+\sqrt{%
(Q_{3}+l_{3})^{2}+l_{\bot }^{2}}-i0.
\end{equation}
In the upper half-plane, instead, there are only two poles, related to the
gluon and the up propagator: 
\begin{equation}
l_{0}=-|\overrightarrow{l}|+i0,\qquad l_{0}=-Q_{0}-\sqrt{%
(Q_{3}+l_{3})^{2}+l_{\bot }^{2}}+i0  \label{upper}
\end{equation}
The poles in (\ref{upper}) are conventionally related to a propagation that
goes backward in time; they will therefore be called the antiparticle poles.
We close for simplicity the integration contour in the upper half-plane and
we have two residue contributions related to the antigluon pole ad the anti-$%
up$ pole respectively (see fig.4): 
\begin{equation}
C_{s}=C_{g}+C_{q}.
\end{equation}
The antigluon and the anti-$up$ contributions are given respectively by 
\begin{eqnarray}
C_{g} &=&\frac{2}{\pi }v\cdot Q\int d^{3}l~\ \left. \ \frac{1}{l_{0}-|%
\overrightarrow{l}|+i0}\,\frac{1}{v\cdot l+i0}\frac{1}{Q^{2}+2l\cdot Q+i0}%
\right| _{l_{0}=-|\overrightarrow{l}|+i0}  \label{3dim} \\
C_{q} &=&\ \frac{2}{\pi }v\cdot Q\int d^{3}l\ \left. \frac{1}{l_{0}+Q_{0}-|%
\overrightarrow{l}+\overrightarrow{Q}|+i0}\frac{1}{v\cdot l+i0}\frac{1}{%
l^{2}+i0}\right| _{l_{0}=-Q_{0}-\sqrt{(Q_{3}+l_{3})^{2}+l_{\bot }^{2}}+i0} 
\nonumber
\end{eqnarray}
As we see by power counting, $C_{g}$ and $C_{q}$ are UV-divergent and it is
therefore necessary to introduce an ultraviolet regularization to treat them
separately. 

The light-cone function factorizes after this second step as

\begin{equation}
F(k_{+})^{QCD}=\frac{1}{-k_{+}+i0}\left[ 1+a\,C_{h}\right] \left[ 1+a\,C_{q}%
\right] \left[ 1+a\,C_{g}\right] ,  \label{Fgrande1}
\end{equation}
i.e. as a product of three factors.

\subsubsection{Wilson line representation}

\bigskip

Before explicitly computing these 3-dimensional integrals, let us represent
them as 4-dimensional ones, i.e. as one-loop integrals of a properly chosen
field theory: 
\begin{equation}
C_{g}\equiv -iv\cdot Q\int \frac{d^{4}l}{\pi ^{2}}\ \frac{1}{Q^{2}+2l\cdot
Q+i0}~\ \frac{1}{v\cdot l+i0}~\ \frac{1}{l^{2}+i0}
\end{equation}
and 
\begin{equation}
C_{q}\equiv iv\cdot Q\int \frac{d^{4}l}{\pi ^{2}}\ \frac{1}{(l+Q)^{2}+i0}~\ 
\frac{1}{v\cdot l+i0}~\ \frac{1}{Q^{2}+2l\cdot Q+i0}  \label{cqrep}
\end{equation}
The proof of the above equations is just by integration over $l_{0}$:
closing the integration contour in the upper half-plane of $l_{0}$, we
enclose a single pole, whose residue gives the 3-dimensional integrals in
eqs.~(\ref{3dim}); $C_{g}$ and $C_{q}$ involve one full - i.e. quadratic -
propagator and two eikonal - i.e. linear - propagators. It is easy to check
that the algebraic sum of $C_{g}$ and $C_{q}$ in the above expressions
reproduces the integral $C_{s}$ defined in eq. (\ref{defC0}). Introducing
the variable $k_{+}$, the integral $C_{g}$ can be written in the
``familiar'' form 
\begin{equation}
C_{g}(k_{+})\equiv \frac{-i}{2}\int \frac{d^{4}l}{\pi ^{2}}\ \frac{1}{%
-k_{+}+l\cdot n+i0}~\ \frac{1}{v\cdot l+i0}~\ \frac{1}{l^{2}+i0}.
\label{feyngauge}
\end{equation}
The geometrical interpretation is the following: $C_{g}(k_{+})$ is the
one-loop correction to a vertex composed of an on-shell Wilson line along
the time axis, and a Wilson line along the direction $n$ off-shell by $k_{+}$%
. Note that 
\begin{equation}
l\cdot n=l_{+}+\frac{n^{2}}{4}l_{-},
\end{equation}
so that eq. (\ref{feyngauge}) represents the vertex correction in Feynman
gauge to the function 
\begin{equation}
F(k_{+})_{n^{2}\neq 0}\equiv \langle B(v)\mid h_{v}^{\dagger }\frac{1}{%
iD_{+}+in^{2}/4D_{-}-k_{+}+i0}h_{v}\mid B(v)\rangle .
\end{equation}
The imaginary part of the above function equals $-1/\pi $ times the shape
function off the light-cone, $n^{2}\neq 0$: 
\begin{equation}
f(k_{+})_{n^{2}\neq 0}\equiv \langle B(v)\mid h_{v}^{\dagger }\,\delta
\left( k_{+}-iD_{+}-in^{2}/4D_{-}\right) \,h_{v}\mid B(v)\rangle .
\end{equation}
In the limit $n^{2}\rightarrow 0$ (see section \ref{nuova}) we recover the
correction to the light-cone function $F(k_{+})$, which was computed in
ref.~ \cite{kg, mn, noi}.

We can give a similar description for $C_{q}$. The 4-dimensional
representation for $C_{q}$ involves an on-shell Wilson line along the time
direction, a Wilson line along the direction $n$ off-shell by $k_{+}$, and a
light quark propagator with momentum $l+Q$ 
\begin{equation}
C_{q}\equiv \frac{i}{2}\int \frac{d^{4}l}{\pi ^{2}}\ \frac{1}{-k_{+}+l\cdot
n+i0}~\ \frac{1}{v\cdot l+i0}\,\,\frac{1}{(l+Q)^{2}+i0}.
\end{equation}
Note that the expressions for $C_{g}$ and $C_{q}$ are very similar: they
differ by an overall sign and by the replacement in the quadratic propagator
of $l\rightarrow l+Q$. For future reference, let us note that the latter
shift involves only the zero and the third components of $l$, not the
transverse ones.

In fig.5 the decomposition of $C_{s}$ into $C_{g}$ and $C_{q}$ is
represented.

\subsubsection{Space Momenta Cutoff}

\label{space}

We consider the bare theory with the regularization introduced in ref. \cite
{noi}: a sharp cutoff on the spatial loop momenta $\Lambda _{S}$ (see next
section). Integrating $C_{g}$ over $l_{0}$ by closing the integration
contour upward and integrating over the azimuthal angle, we obtain

\begin{equation}
C_{g}=-\int_{0}^{\Lambda _{S}}dl\int_{-1}^{1}d\cos \theta \frac{1}{%
k_{+}+l\left( 1-\cos \theta \right) +n^{2}l/4\left( 1+\cos \theta \right) },
\label{starting}
\end{equation}
where we have used the definition of $Q^{\mu }$ in eq.~(\ref{defQbig}) and $%
l\equiv |\overrightarrow{l}|$. Integrating over the polar angle, we obtain 
\begin{equation}
C_{g}=-\int_{0}^{2\Lambda _{S}}\frac{dl}{l}\log \left[ \frac{k_{+}+l}{%
k_{+}+n^{2}\,l/4}\right] .  \label{basic1}
\end{equation}
Assuming a cutoff much larger than any physical scale in the process, i.e.%
\footnote{%
This is done consistently with the relation (\ref{largecut}), in which we
have taken a large cutoff for the computation of $C_{s}$.} 
\begin{equation}
\Lambda _{S}\gg E_{X},  \label{realcon}
\end{equation}
we obtain in DLA \footnote{%
The last member in eq.~(\ref{firstres}) is an artificial absorptive part
that cancels against an opposite one in $C_{q}$ (see eq.~(\ref{basic2}).} 
\begin{equation}
C_{g}=-\frac{1}{2}\log ^{2}\frac{E_{X}}{k_{+}-i0}-\log \frac{\Lambda _{S}}{%
E_{X}}\log \frac{E_{X}}{-k_{+}+i0}.  \label{firstres}
\end{equation}
Three scales enter in $C_{g}$: $k_{+},\,\Lambda _{S}$ and $E_{X}.$ The
appearance of $k_{+}$ and the cutoff $\Lambda _{S}$ was expected, because
these two quantities represent the infrared and the ultraviolet scale,
respectively. The noticeable fact is that also the hadronic energy $E_{X}$
makes its appearance. $C_{g}$ contains a double-logarithm of the infrared
kinematical scale $k_{+}$ (related to the overlap of the soft and the
collinear region, which extends up to $\Lambda _{S}$); it also contains a
single logarithm of the cutoff. The appearance of the hadronic energy $E_{X}$
comes from the necessity of a third mass scale for the function $C_{g}$,
which behaves like $\log ^{2}k_{+}$ in $k_{+}$ and like $\log \Lambda _{S}$
in $\Lambda _{S}.$

When $l\gg E_{X}$ the integrand behaves as 
\begin{equation}
-\frac{1}{l}\log \left[ \frac{k_{+}+l}{k_{+}+n^{2}\,l/4}\right] \sim \frac{1%
}{l}\log \frac{n^{2}}{4}  \label{singlelo}
\end{equation}
and produces a single-logarithmic ultraviolet divergence. As eq.~(\ref
{singlelo}) clearly indicates, $n^{2}\neq 0$ regulates the collinear or
light-cone singularity: up to now we have indeed taken kinematics into
account exactly together with a large cutoff. It is interesting to note that
if we take a cutoff much smaller than the hadronic energy (as we will do in
the ``final'' low-energy effective theory), 
\begin{equation}
\Lambda _{ET}\,\ll \,E_{X},  \label{smallcutoff}
\end{equation}
we have 
\begin{equation}
n^{2}\,l\,\leq \,\frac{m_{X}^{2}}{E_{X}^{2}}\,\,\Lambda _{ET}\,\ll \,k_{+},
\end{equation}
and $C_{g}$ simplifies in 
\begin{equation}
C_{g}\left( \Lambda _{ET}\right) \simeq -\int_{0}^{2\Lambda _{ET}}\frac{dl}{l%
}\log \left( \frac{k_{+}+l}{k_{+}}\right) .  \label{onthelight}
\end{equation}
The quantity $n^{2}$ does not enter anymore and the integrand is the same as
that with the approximate light-cone kinematics $n^{2}=0$, i.e. with $n$
replaced by $\overline{n}$. The physical explanation is that soft gluons are
not able to distinguish between the two slightly different directions $n$
and $\overline{n}$. Formally, with the small cutoff (\ref{smallcutoff}) we
can take the limit (\ref{Sudakov}) inside the integral. In other words, in
the low-energy effective theory, we effectively are always in the light-cone
limit.\ For $l\gg k_{+}$, the integrand in eq.~(\ref{onthelight}) has the
asymptotic behaviour 
\begin{equation}
\frac{1}{l}\log \frac{l}{k_{+}},
\end{equation}
implying a double-logarithmic behaviour with respect to $\Lambda _{ET}$ upon
integration over $l$, in contrast with the single logarithmic behaviour of
the integrand in (\ref{singlelo}). These properties will be studied
systematically in the next section, in which we consider the effective
theory on the light-cone.

For the computation of $C_{q}$, it is convenient to first make the shift $%
l\rightarrow l-Q$ in the expression of $C_{q},$\ eq.~(\ref{cqrep}), and then
to compute the residue of the light quark pole at $l_{0}=-|\overrightarrow{l}%
|+i0$: this is legitimate if condition (\ref{realcon}) holds. We find 
\begin{equation}
C_{q}=\log \frac{\Lambda _{S}}{E_{X}}\log \frac{E_{X}}{-k_{+}+i0} \quad
\quad \quad (\Lambda_S \gg E_X).  \label{basic2}
\end{equation}
The three scales appearing in $C_{g}$ do appear also in $C_{q}$. We note
that $C_{q}$ contains a single logarithm of $k_{+}$, i.e. it is subleading
by one logarithm in the infrared counting with respect to $C_{g}$. It has a
single-logarithmic UV divergence.

If we compute the integral in eq.~(\ref{cqrep}) with a small cutoff (\ref
{smallcutoff}), we do not find any infrared logarithm, in contrast with what
happens instead with $C_{g}$. Thus the logarithmic contributions to $C_{q}$
come from high-energy gluons and that is an indication that $C_{q}$, unlike $%
C_{g}$, is short-distance dominated.

One can check that the correct value of $C_{s}$ is reproduced by summing $%
C_{g}$ and $C_{q}$; in particular, UV divergences cancel.

At the level of logarithmic accuracy, we can replace the strong inequality (%
\ref{realcon}) with a weaker one: 
\begin{equation}
\Lambda _{S}\geq E_{X}.
\end{equation}
Setting in particular 
\begin{equation}
\Lambda _{S}=E_{X},
\end{equation}
the expressions for the gluon and quark-pole residue read 
\begin{eqnarray}
C_{g}\left( \Lambda _{S}=E_{X}\right) &=&-\frac{1}{2}\log ^{2}\frac{E_{X}}{%
k_{+}-i\epsilon }=C_{s},  \nonumber \\
C_{q}\left( \Lambda _{S}=E_{X}\right) &=&0,
\end{eqnarray}
i.e. the gluon-pole term gives the whole contribution while the quark-pole
factor vanishes. The term $C_{q}$ therefore has the role of correcting $%
C_{g} $ when $\Lambda _{S}\neq E_{X}.$

Since ultraviolet singularities are single-logarithmic for a large cutoff
(eqs. (\ref{firstres}) and (\ref{basic2})), other regularizations such as DR
give similar results. In other words, the regularization scheme dependence
is the usual one: the logarithmic term in the one-loop amplitude is scheme
independent while the finite part is scheme dependent.

\section{ The effective theory on the light-cone, the LEET}

\label{nuova}

In full QCD infrared singularities are regulated by the unique quantity $%
m_{X}^{2}\neq 0$. In the effective theory, the light quark propagator is
replaced by an eikonal one 
\begin{equation}
\frac{1}{\left( l+Q\right) ^{2}+i0}=\frac{1}{Q^{2}+2l\cdot Q+l^{2}+i0}%
\,\,\rightarrow \,\,\frac{1}{Q^{2}+2l\cdot Q+i0}.
\end{equation}
In the expression on the right-hand side $l^{2}$ has been neglected and as a
consequence $Q_{\mu }$ enters in two distinct and independent ways: its
square $Q^{2}$ represents the virtuality of the eikonal line at $l_{\mu }=0$%
, while its components $Q^{\mu }$ are the coefficients of the linear
combination of the loop-momentum components $l_{\mu }$ in the term $%
2\,l_{\mu }\,Q^{\mu }$. Unlike full QCD, $Q^{2}$ and $Q^{\mu }$ can be
considered as independent quantities in the effective theory. We may ask
ourselves what happens if we take the limit $Q^{2}\rightarrow 0$ inside the
term $2l\cdot Q\rightarrow 2l\cdot \overline{Q}$ while keeping 
$Q^{2}\rightarrow {\rm const}\neq 0$, i.e. if we make the replacement 
\begin{equation}
\frac{1}{Q^{2}+2l\cdot Q+i0}\,\rightarrow \,\frac{1}{Q^{2}+2l\cdot \overline{%
Q}+i0}.
\end{equation}
In the usual notation, the above replacement reads 
\begin{equation}
\frac{1}{-k_{+}+n\cdot l+i0}\,\rightarrow \,\frac{1}{-k_{+}+\overline{n}%
\cdot l+i0},
\end{equation}
corresponding to the limit 
\begin{equation}
n^{2}\rightarrow 0,\quad k_{+}\rightarrow {\rm const}\neq 0.
\label{lightcone}
\end{equation}
This means that we are considering an eikonal propagator that lies exactly
on the light-cone, with collinear singularities regulated now by $k_{+}\neq 0
$ only, instead of by $n^{2}\neq 0$ \cite{renshape, cattren, firstleet,
instableet}. As we saw before, the limit $n^{2}\rightarrow 0$ is
``invisible'' with a small cutoff, simply because the integrand does not
depend on $n^{2}$ in this case. We now want to see what happens in the limit
(\ref{lightcone}) with a {\it large} cutoff. To perform IR factorization in
the light-cone case, it is convenient to start from the original QCD
amplitude $C_{s}$ in which we make the replacement 
\begin{equation}
\frac{1}{Q^{2}+2l\cdot Q+l^{2}+i0}\,\,\rightarrow \,\,\frac{1}{Q^{2}+2l\cdot 
\overline{Q}+l^{2}+i0},  \label{basappr}
\end{equation}
to obtain 
\begin{equation}
\,C_{s}\,\rightarrow \,\overline{C}_{s}\equiv -iv\cdot Q\int \frac{d^{4}l}{%
\pi ^{2}}\ \frac{1}{Q^{2}+2l\cdot \overline{Q}+l^{2}+i0}~\ \frac{1}{v\cdot
l+i0}~\ \frac{1}{l^{2}+i0}.
\end{equation}
It is straightforward to check that $C_{s}$ and $\overline{C}_{s}$ coincide
at the DLA level, i.e. that the approximation (\ref{basappr}) \ preserves
the infrared structure, 
\begin{equation}
\overline{C}_{s}\,=\,C_{s}\,=\,-\frac{1}{2}\log ^{2}(-n^{2}-i0)\qquad \quad
(DLA).
\end{equation}
The gluon and the quark pole contributions are given in the light-cone limit
by 
\begin{eqnarray}
C_{g,n^{2}=0} &\equiv &-iv\cdot Q\int \frac{d^{4}l}{\pi ^{2}}\ \frac{1}{%
Q^{2}+2l\cdot \overline{Q}+i0}~\ \frac{1}{v\cdot l+i0}~\ \frac{1}{l^{2}+i0}
\\
C_{q,n^{2}=0} &\equiv &iv\cdot Q\int \frac{d^{4}l}{\pi ^{2}}\ \frac{1}{%
Q^{2}+2l\cdot \overline{Q}+l^{2}+i0}~\ \frac{1}{v\cdot l+i0}~\ \frac{1}{%
Q^{2}+2l\cdot \overline{Q}+i0}.  \nonumber
\end{eqnarray}
The above terms are usually called ``soft'' and ``jet'' (or ``collinear'')
factor respectively, even though we believe that this terminology can be
rather misleading, as the redistribution of 
double logarithmic contributions
in $C_{g}$ and $C_{q}$ is substantially dependent on the regularization. We
will show later that it is possible, within a specific class of
regularization schemes, to confine all the 
double logarithmic contributions in $C_{g}.$ It
is immediate to check that the two above integrands sum up to the
integrand
of $\overline{C}_{s}$. Making the shift $l\rightarrow l-\overline{Q}$, the
quark factor can also be written as 
\begin{equation}
C_{q,n^{2}=0}\equiv iv\cdot Q\int \frac{d^{4}l}{\pi ^{2}}\ \frac{1}{%
l^{2}+Q^{2}+i0}~\ \frac{1}{v\cdot l-v\cdot \overline{Q}+i0}~\ \frac{1}{%
Q^{2}+2l\cdot \overline{Q}+i0}.
\end{equation}

\subsection{\protect\bigskip Regularization Effects}

The decomposition of \ $\overline{C}_{s}$ into $C_{g,n^{2}=0}$ and $%
C_{q,n^{2}=0}$ is strongly dependent on the regularization scheme adopted,
as a consequence of the fact that double-infrared logarithms are promoted to
double-ultraviolet logarithms with the splitting. We will see that there are
substantial regularization scheme effects, even for the leading DLA terms.
Two different classes of regularizations are considered. To the first class
belongs the regularization considered in ref.~\cite{noi}: a sharp cutoff on
the spatial loop momenta 
\[
|\overrightarrow{l}|<\Lambda _{S},
\]
and a loop energy on the entire real axis, 
\begin{equation}
-\infty <l_{0}<\infty .  \label{3b}
\end{equation}
That means, roughly speaking, a discrete space and a continuous time. We
believe that this regularization gives the same double-logarithm as the
ordinary lattice regularization - the Wilson action \cite{mesolo}. In the
latter case all the components of the loop $4$-momentum are cutoff, not only
the spatial ones 
\begin{equation}
|l^{\mu }|<\Lambda _{4}\equiv \frac{\pi }{a},  \label{lat4}
\end{equation}
where $a$ is the lattice spacing. The physical reason for the equality of
the double-logarithmic coefficients in the regularizations (\ref{3b}) and (%
\ref{lat4}) is the following. Soft and collinear logarithms are both related
to quasi-real gluon configurations, for which 
\begin{equation}
l_{0}\sim |\overrightarrow{l}|.
\end{equation}
Cutting off the spatial momenta therefore should cut off also the relevant
energies as far as soft and collinear singularities are concerned.

As a representative of the second class of UV regularizations, consider a
sharp cutoff on the transverse momenta (the $x$--$y$ plane): 
\begin{equation}
|\overrightarrow{l}_{\perp }|<\Lambda _{\perp },\quad {\rm while}\quad
-\infty <l_{+},\,\,l_{-}<\infty .
\end{equation}
This regularization is ``effective'', i.e. it is sufficient to cut on the
transverse momenta to render the integrals finite. To this class of
regularizations belongs the Dimensional Regularization (DR), in which most
of the effective field theory computations have been performed. Let us treat
the two cases in turn.

\subsubsection{Space Momenta Cutoff}

An explicit computation of the gluon and the quark pole contributions on the
light-cone in the $\Lambda _{S}$-regularization gives 
\begin{eqnarray}
C_{g,\,n^{2}=0} &=&-\frac{1}{2}\log ^{2}\frac{\Lambda _{S}}{k_{+}-i0} 
\nonumber \\
C_{q,\,n^{2}=0} &=&\frac{1}{2}\log ^{2}\frac{\Lambda _{S}}{k_{+}-i0}-\frac{1%
}{2}\log ^{2}\frac{E_{X}}{k_{+}-i0}.  \label{firstres2}
\end{eqnarray}
The behaviour with respect to $k_{+}$ is the same as in the case $n^{2}\neq
0.$ Ultraviolet divergences are now more severe than in the case $n^{2}\neq
0 $, being of double-logarithmic kind. However, the sum is again the correct
one 
\begin{equation}
C_{g,\,n^{2}=0}+C_{q,\,n^{2}=0}=C_{s}.  \label{somma_1}
\end{equation}
In other words, the transition to the light-cone theory implies a
rearrangement of the ultraviolet structure, but the physical observable, $%
C_{s}$, is unchanged.

\subsubsection{ Transverse momenta cutoff}

The factor $C_{g}$ is better computed in this case by introducing light-cone
coordinates: 
\begin{equation}
l_{+}=l_{0}+l_{3},\quad l_{-}=l_{0}-l_{3}.
\end{equation}
Integrating over $l_{-}$ by closing the integration contour upward and over
the transverse momentum, we obtain 
\begin{equation}
C_{g}=-2\int_{0}^{\infty }\frac{dy\,y}{1+y^{2}}\,\frac{1}{1+n^{2}y^{2}/4}%
\log \left[ 1+\frac{\Lambda _{\perp }}{k_{+}y}\left( 1+\frac{n^{2}y^{2}}{4}%
\right) \right] .
\end{equation}
Performing the final integration in the case $n^{2}=0$, we obtain 
\begin{equation}
C_{g,\,n^{2}=0}\left( \Lambda _{\bot }\right) =-\log ^{2}\frac{\Lambda
_{\perp }}{k_{+}-i0}.  \label{cg}
\end{equation}
For the quark-pole factor $C_{q},$ the integration over $l_{-}$ gives 
\begin{equation}
C_{q}=\int_{0}^{\infty }dx\,x\int_{0}^{\Lambda _{\perp }^{2}}dl_{\perp
}^{2}\,\frac{1}{x^{2}l_{\perp }^{2}+2E_{X}\,x+1}\,\,\frac{1}{k_{+}\,x+1-i0}.
\end{equation}
Integrating over $l_{\perp }$ we obtain, in the light-cone limit: 
\begin{equation}
C_{q}=\int_{0}^{\infty }\frac{dx}{x}\,\frac{1}{1-n^{2}\,x/4\,-i0}\,\log 
\frac{\overline{\Lambda }_{\perp }^{2}\,x^{2}+x+1}{1+x},
\end{equation}
with\footnote{$\,n^{2}$ in the above formula has to be interpreted as $%
-2k_{+}/E_{X}$.} 
\begin{equation}
\overline{\Lambda }_{\perp }\equiv \frac{\Lambda _{\perp }}{2E_{X}}.
\end{equation}
The above integral has two double-logarithmic regions for $\Lambda _{\perp
}\gg E_{X},$ 
\begin{equation}
(1):\,1\ll x\ll \frac{4}{n^{2}},\quad \quad (2):\,\frac{1}{\overline{\Lambda 
}_{\perp }}\ll x\ll 1.
\end{equation}
Performing the integration in the two regions, we find 
\begin{equation}
C_{q,\,n^{2}=0}\left( \Lambda _{\bot }\right) =\frac{1}{2}\log ^{2}\frac{%
\Lambda _{\perp }^{2}}{E_{X}\left( k_{+}\,-i0\right) }-\log ^{2}\frac{%
\Lambda _{\perp }}{E_{X}}\quad \quad \quad (\Lambda _{\bot }\gg E_{X}).
\label{cq}
\end{equation}
The first double-logarithm on the right-hand side is related to region $(1)$%
, the second one to region $(2)$. For a smaller UV cutoff, we obtain
instead: 
\begin{equation}
C_{q,\,n^{2}=0}\left( \Lambda _{\bot }\right) =\frac{1}{2}\log ^{2}\frac{%
\Lambda _{\perp }^{2}}{E_{X}\left( k_{+}\,-i0\right) },\quad \quad \quad
\left( E_{X}\,|k_{+}|\ll \Lambda _{\perp }^{2}\ll E_{X}^{2}\right) .
\end{equation}
Finally, for $\Lambda _{QCD}^{2}\ll \Lambda _{\perp }^{2}\ll E_{X}\,|k_{+}|$%
, the integral $C_{q}$ vanishes in DLA. 

\subsubsection{ Comments }

Let us comment on the results (\ref
{cg}) and (\ref{cq}). As with the 3-momentum regularization, $C_{g}$ and $%
C_{q}$ have double-logarithmic UV divergences, again a consequence of the
light-cone limit $n^{2}=0$. The most important point, however, is that $%
C_{g} $ has an additional factor of 2 with respect to the spatial cutoff
case in the coefficient of the double-logarithm of the infrared scale, $\log
^{2}k_{+}$ (cf. eqs.~(\ref{firstres}) and (\ref{cg})). With the $\Lambda
_{S} $ regularization, $C_{q}$ has no $\log ^{2}k_{+}$ term, while with the $%
\Lambda _{\perp }$ regularization it does. The same double logarithm is
obtained in the sum $C_{s}$ in both regularizations. In general, the
appearance of $\log ^{2}k_{+}$ in $C_{q}(\Lambda _{\perp })$ implies that,
with the $\Lambda _{\perp }$ regularization, $C_{q}$ does not describe only
collinear contributions but also soft ones \footnote{%
The double logarithm necessarily comes from the overlap.}. We interpret this
fact by saying that 
 the shape function, in general, does not have any physical meaning,
but it just represents the gluon-pole contribution to a physical process:
that result is, as far as we know, new. One generally attaches to the shape
function a physical meaning - related to the Fermi motion; thus, to
understand what is happening, we have to start again from the beginning. The
shape function is obtained from the original QCD tensor $W_{\mu \nu }$
considering the infrared limit of small momenta compared with the hadronic
energy: 
\begin{equation}
|l_{\mu }|\,\ll \,E_{X}.  \label{small}
\end{equation}
The tree-level rate in the ET equals the QCD one by construction. However,
in loops, the condition (\ref{small}) is not guaranteed: its validity
depends on the regularization scheme adopted. If we cut all the
loop-momentum components with a hard cutoff much smaller than the hard
scale, 
\begin{equation}
|l_{\mu }|\,\leq \,\Lambda _{UV}\,\ll \,E_{X},
\end{equation}
then the condition (\ref{small}) is still valid at the loop level. As a
consequence, we expect that the leading, double-logarithmic term of the ET
shape function will match the QCD one. That is indeed what happens with the
spatial momentum regularization, as we have seen explicitly. On the other
hand, when one uses a regularization such as DR or $\Lambda _{\perp }$, the
equality of the double-logs is no longer guaranteed, and indeed it does not
occur in $\Lambda _{\perp }$-regularization, as we have seen explicitly.
This is because the longitudinal momentum of the gluon $l_{z}$, or
equivalently its energy $\epsilon $, can become arbitrarily large. For the
latter regularizations, even for the double-logarithm, one has to come back
to the original QCD loop diagram and perform factorization into a factor $%
C_{g}$ and a factor $C_{q}$, as we have shown in detail. In ref.~\cite{noi}
it was shown that the factor of 2 in the $\log ^{2}k_{+}$ term in DR is a
regularization effect, i.e. it can be removed by going to a non-minimal
dimensional scheme. We explicitly see, with the similar $\Lambda _{\perp }$
regularization, that by including $C_{q}$ the scheme-dependence
automatically disappears. The origin of the additional factor of 2 in the
transverse-momentum regularization is related to the occurrence of a second
double-logarithmic region for $|l_{z}|\gg \Lambda_{\perp} $ 
(very large  rapidity).

Finally, as already noted, let us observe that in the case $n^{2}\neq 0$ we
expect the transverse momentum cutoff to give double-logarithmic results
similar to those from the space momentum cutoff. That is because $n^{2}\neq
0 $ cuts the collinear emission at infinite rapidity.

\section{The shape function in the 
low-energy effective theory}

\label{effective_theory}

With the $\Lambda _{\perp }$ regularization, double logarithms
 are contained in $C_{g}$ as well as in $C_{q}$. Since
we want to confine double logarithmic effects inside the shape function only, let
us consider from now on the $\Lambda _{S}$ regularization only. The factor $%
C_{q}$ is short-distance dominated in the latter regularization, so it is
computed once and for all in perturbation theory and ``leaves the game''.

Let us therefore return to formula (\ref{starting}) for $C_{g}$. Calling $%
\epsilon =|\overrightarrow{l}|,$ and $t=\theta ^{2},$ $C_{g}$ can be written
as 
\begin{eqnarray}
C_{g}\left( \Lambda _{S},k_{+}\right) &\simeq &-\int_{0}^{\Lambda
_{S}}d\epsilon \int_{0}^{1}dt\frac{1}{2k_{+}+\epsilon \,t}  \nonumber \\
&\simeq &-\int_{0}^{\Lambda _{S}}\frac{d\epsilon }{\epsilon }\int_{0}^{1}%
\frac{dt}{t}\theta \left( \epsilon t-k_{+}\right)  \nonumber \\
&\simeq &-\frac{1}{2}\log ^{2}\frac{\Lambda _{S}}{k_{+}},
\end{eqnarray}
where we have assumed $\Lambda _{S}\lesssim O\left( E_{X}\right) $ and we
have used the approximation $1/\left( 2k_{+}+\epsilon \,t\right) \simeq
\theta \left( \epsilon t-2k_{+}\right) /\left( \epsilon \,t\right) $, which
is valid in DLA. This form helps visualizing the origin of the double
logarithm. We see that contributions come from soft regions, where $\epsilon
\sim O\left( k_{+}\right) $, as well as from hard regions, where $\epsilon
\sim O\left( \Lambda _{S}\right) $. In order to separate them, the simplest
way is to introduce another UV cutoff $\Lambda _{ET}$, this time well below
the hadronic energy $E_{X}$ (the hard scale of the process), such as 
\begin{equation}
k_{+}\ll \Lambda _{ET}\ll \Lambda _{S}.
\end{equation}
We can write 
\begin{equation}
C_{g}\left( \Lambda _{S},k_{+}\right) =\delta Z\left( \Lambda _{S},\Lambda
_{ET},k_{+}\right) +\delta \bar{F}^{ET}\left( \Lambda _{ET},k_{+}\right) ,
\end{equation}
where 
\begin{eqnarray}
\delta Z\left( \Lambda _{S},\Lambda _{ET},k_{+}\right) &\equiv&
-\int_{\Lambda _{ET}}^{\Lambda _{S}}\frac{d\epsilon }{\epsilon }\int_{0}^{1}%
\frac{dt}{t}\theta \left( \epsilon t-k_{+}\right)  \nonumber \\
&=& -\frac{1}{2} \log^2 \frac{\Lambda_S}{k_{+}}+ \frac{1}{2} \log^2 \frac{%
\Lambda_{ET}}{k_{+}}  \label{laststep}
\end{eqnarray}
is a coefficient function and $\delta \bar{F}^{ET}\left( \Lambda
_{ET},k_{+}\right) $ is the one-loop contribution to the light-cone function 
$\delta F^{ET}$, multiplied by the propagator: $\delta F^{ET}= \delta \bar{F}%
^{ET}/(-k_{+}+i0) $, as defined in eq.~(\ref{defFhqet}), 
\begin{eqnarray}
\delta \bar{F}^{ET}\left( \Lambda _{ET},k_{+}\right) &\equiv&
-\int_{0}^{\Lambda _{ET}}\frac{d\epsilon }{\epsilon }\int_{0}^{1}\frac{dt}{t}%
\theta \left( \epsilon t-k_{+}\right)  \nonumber \\
&=&-\frac{1}{2}\log ^{2}\left( \frac{\Lambda _{ET}}{k_{+} }\right) .
\label{tof}
\end{eqnarray}
Note that $\delta \bar{F}^{ET}$ depends only on the two scales $k_{+}$ and $%
\Lambda _{ET}$. This is in line with the idea of a simple low-energy
effective theory, which describes infrared phenomena characterized by the
scale $k_{+}$, apart from the UV cutoff that enters through loop effects.

We assume that long-distance effects can be traced by the growth of the
coupling constant in the proximity of the Landau pole, and that the coupling
constant must be evaluated at the transverse momentum squared~\cite{ven}: 
\begin{equation}
\alpha _{S}\rightarrow \alpha _{S}\left( k_{\perp }^{2}\right) ,
\end{equation}
where 
\begin{equation}
k_{\perp }^{2}\cong \epsilon ^{2}t.
\end{equation}
From the expression of $\delta \,Z$ we see that transverse momenta have a
lower bound given by 
\begin{equation}
l_{\perp }^{2}>l_{\perp }^{2}{}_{\min }=\Lambda _{ET}\,k_{+}.  \label{impor}
\end{equation}
According to our criteria, non-perturbative effects are absent from $Z$ as
long as 
\begin{equation}
l_{\perp }^{2}{}_{\min }\,\gg \,\Lambda _{QCD}^{2}.  \label{shortdis}
\end{equation}
According to eq.~(\ref{impor}), this occurs when $k_{+}$ is non-vanishing,
as it is for example if 
\begin{equation}
k_{+}\sim O\left( \Lambda _{QCD}\right) ,
\end{equation}
as expected from Fermi motion (since $\Lambda _{ET}\gg \Lambda _{QCD}$)$.$
However, by taking the imaginary part of $T_{\mu \nu }$ to obtain $W_{\mu
\nu }$, i.e. the rate, the product of factors is converted into a
convolution over $k_{+}$ and the point $k_{+}=0$ is included in the
integration range. This implies that transverse momenta down to zero
contribute to the coefficient function in $W_{\mu \nu }$ , i.e. that
factorization of short- and long-distance effects breaks down at this point.
The breakdown is related to the fact that we are cutting the energies of the
gluons, but not the emission angles, which can go down to zero, implying the
vanishing of the transverse momenta. That is one of the most important
outcomes of our analysis. However, we believe that these long-distance
contributions are suppressed. Let us present a qualitative argument. As we
can see from inequalities (\ref{impor}) and (\ref{shortdis}), transverse
momenta of the order of the hadronic scale occur in $Z$ for a very small
slice of values of $k_{+}$, 
\begin{equation}
k_{+}\lesssim \frac{\Lambda _{QCD}^{2}}{\Lambda _{ET}}\ll \Lambda _{QCD}.
\end{equation}
If the integrand is not singular in this small slice, as it is natural to
assume, it gives a reasonally small fraction of the total. Note that the
usual factorization of mass singularities is instead\ ``exact''. If we
consider for example the moments of DIS cross-section, factorization
involves a splitting of the long- and short-distance contributions of the
form 
\begin{eqnarray}
M_{N}\left( Q^{2}\right) &=&\int_{0}^{1}dx_{B}\,x_{B}^{N-1}\,\sigma
_{DIS}\left( x_{B},Q^{2}\right) \\
&=&1+\gamma _{N}\alpha _{S}\int_{m^{2}}^{Q^{2}}\frac{dl_{\perp }^{2}}{%
l_{\perp }^{2}}=\left( 1+\gamma _{N}\,\alpha _{S}\int_{\Lambda ^{2}}^{Q^{2}}%
\frac{dl_{\perp }^{2}}{l_{\perp }^{2}}\right) \left( 1+\gamma _{N}\,\alpha
_{S}\int_{m^{2}}^{\Lambda ^{2}}\frac{dl_{\perp }^{2}}{l_{\perp }^{2}}\right)
,  \nonumber
\end{eqnarray}
where $m$ is the mass of a light quark.

After the last step (\ref{laststep}), the forward hadronic tensor takes the
final form 
\begin{eqnarray}
T_{\mu \nu }^{QCD} &=&\frac{s_{\mu \nu }}{2v\cdot Q}\,F(k_{+})^{QCD} \\
&=&\frac{s_{\mu \nu }}{2v\cdot Q}\,\frac{1}{-k_{+}+i0}\left[ 1+a\,C_{h}%
\right] \left[ 1+a\,C_{q}\right] \left[ 1+a\,\delta Z\right] \left[
1+a\,\delta \bar{F}^{ET}\right] ,  \nonumber  \label{last_step_1}
\end{eqnarray}
where the various factors have been introduced in eqs.~(\ref{Tforward}), (%
\ref{Fgrande}), (\ref{LCfunction}) and (\ref{Fgrande1}). Taking the
imaginary (absorptive) part, according to the optical theorem (\ref{optical}%
), we have for $W_{\mu \nu }$ the multiple convolution 
\begin{eqnarray}
W_{\mu \nu } &=&\frac{s_{\mu \nu }}{2v\cdot Q}\int
dk_{1}\,dk_{2}\,dk_{3}\,dk_{4}\,\delta \left(
k_{+}-k_{1}-k_{2}-k_{3}-k_{4}\right)  \nonumber \\
&&\left[ \delta \left( k_{1}\right) +a\,c_{h}\left( k_{1}\right) \right] %
\left[ \delta \left( k_{2}\right) +a\,c_{q}\left( k_{2}\right) \right] 
\nonumber \\
&&\left[ \delta \left( k_{3}\right) +a\,\delta z\left( k_{3}\right) \right] %
\left[ \delta \left( k_{4}\right) +a\delta f^{ET}\left( k_{4}\right) \right]
,
\end{eqnarray}
where 
\begin{equation}
f^{ET}\left( k_{+},\Lambda _{ET}\right) =\delta \left( k_{+}\right)
+a\,\delta f^{ET}\left( k_{+}\right) +O(a^{2})
\end{equation}
is the shape function, defined in eq.~(\ref{defshape}), for an on-shell
quark ($k_{+}^{\prime }=0$); moreover, we have defined 
\begin{eqnarray}
c_{h}\left( k_{+}\right) &\equiv &-\frac{1}{\pi }{\rm {Im}}\left[ \frac{1}{%
-k_{+}+i0}C_{h}\left( k_{+}-i0\right) \right]  \nonumber \\
&=&\delta \left( k_{+}\right) C_{h}\left( k_{+}\right) -\frac{1}{k_{+}}%
\left( -\frac{1}{\pi }\right) {\rm {Im}}\,C_{h}\left( k_{+}-i0\right)
\end{eqnarray}
and analogously for the other factors\footnote{%
In ref.~\cite{noi}, formula (9) should be replaced by $f^{QCD}(k_{+})=\int
dk_{1}\,dk_{2}\,\,\delta (k_{+}-k_{1}-k_{2})\,(\delta (k_{1})+a\,\delta
z(k_{1}))\,f^{ET}(k_{2})$, where $Z=1+a\,\delta Z$.}. Typically, by taking
the imaginary parts, for double-logarithmic contributions, we have 
\begin{eqnarray}
-\frac{1}{\pi }\,\frac{\log ^{2}\left( k_{+}-i0\right) }{-k_{+}+i0}\,
&\rightarrow &\,\delta \left( k_{+}\right) \log ^{2}(-k_{+})+2\theta \left(
-k_{+}\right) \frac{\log \left( -k_{+}\right) }{-k_{+}}  \nonumber \\
&=&\frac{d}{dk_{+}}\left( -\theta \left( -k_{+}\right) \log ^{2}\left(
-k_{+}\right) \right)
\end{eqnarray}
and for single-logarithmic ones 
\begin{eqnarray}
-\frac{1}{\pi }\,\frac{\log \left( k_{+}-i0\right) }{-k_{+}+i0}\,
&\rightarrow &\,\delta \left( k_{+}\right) \log (-k_{+})+\theta \left(
-k_{+}\right) \frac{1}{-k_{+}}  \nonumber \\
&=&\frac{d}{dk_{+}}\left( -\theta \left( -k_{+}\right) \log \left(
-k_{+}\right) \right) .
\end{eqnarray}
The last members of the above equations have to be interpreted as
distributions. In DLA, according to eq.~(\ref{tof}), $f^{ET}$ up to one loop
reads 
\begin{eqnarray}
f^{ET}\left( k_{+},\Lambda _{ET}\right) &=&\delta \left( k_{+}\right)
+a\,\,\theta \left( -k_{+}\right) \frac{\log \Lambda _{ET}/\left(
-k_{+}\right) }{-k_{+}}-\frac{a}{2}\,\delta \left( k_{+}\right) \log
^{2}\left( \frac{\Lambda _{ET}}{-k_{+}}\right)  \nonumber \\
&=&\delta \left( k_{+}\right) +\frac{a}{2}\frac{d}{dk_{+}}\left( \theta
\left( -k_{+}\right) \log ^{2}\left( \frac{\Lambda _{ET}}{-k_{+}}\right)
\right) .
\end{eqnarray}

\subsection{ Evolution}

Taking a derivative with respect to the logarithm of the cutoff, we obtain 
\begin{eqnarray}
\frac{d\,f\left( k_{+},\Lambda _{ET}\right) }{d\,\log \Lambda _{ET}}
&=&-a\,\delta \left( k_{+}\right) \log \left( \frac{\Lambda _{ET}}{-k_{+}}%
\right) +a\,\,\frac{\theta \left( -k_{+}\right) }{-k_{+}}  \nonumber \\
&=&a\frac{d}{dk_{+}}\left( \theta \left( -k_{+}\right) \log \left( \frac{%
\Lambda _{ET}}{-k_{+}}\right) \right) .
\end{eqnarray}
Comparing the above equation with the evolution equation for the shape
function 
\begin{equation}
\frac{d\,f\left( k_{+},\Lambda _{ET}\right) }{d\,\log \Lambda _{ET}}=\int
dk_{+}^{\prime }\,{\rm K}_{S}\left( k_{+}-k_{+}^{\prime };\Lambda
_{ET}\right) \,f\left( k_{+}^{\prime },\Lambda _{ET}\right) ,
\label{evshape}
\end{equation}
and taking into account that, at lowest order in $\alpha _{S}$, $f\left(
k_{+}^{\prime },\Lambda _{ET}\right) =\delta \left( k_{+}^{\prime }\right) $
holds, we find for the evolution kernel at one loop 
\begin{eqnarray}
{\rm K}_{S}\left( k_{+}-k_{+}^{\prime };\Lambda _{ET}\right) &=&-a\,\delta
\left( k_{+}^{\prime }-k_{+}\right) \log \left( \frac{\Lambda _{ET}}{%
k_{+}^{\prime }-k_{+}}\right) +a\,\,\frac{\theta \left( k_{+}^{\prime
}-k_{+}\right) }{k_{+}^{\prime }-k_{+}}  \nonumber \\
&=&a\frac{d}{dk_{+}}\left( \theta \left( k_{+}^{\prime }-k_{+}\right) \log
\left( \frac{\Lambda _{ET}}{k_{+}^{\prime }-k_{+}}\right) \right)
\label{kspace} \\
&=&a\left[ \frac{\theta \left( k_{+}^{\prime }-k_{+}\right) }{k_{+}^{\prime
}-k_{+}}-\delta \left( k_{+}^{\prime }-k_{+}\right) \int_{0}^{\Lambda _{ET}}%
\frac{d\left( l_{+}^{\prime }-l_{+}\right) }{l_{+}^{\prime }-l_{+}}\right] .
\nonumber
\end{eqnarray}
If we consider the $\Lambda _{\perp }$-regularization, the evolution kernel
for the shape function is instead given by (eq.~(\ref{cg})): 
\begin{eqnarray}
K_{\perp }\left( k_{+}-k_{+}^{\prime };\Lambda _{\perp }\right)
&=&-2a\,\delta \left( k_{+}^{\prime }-k_{+}\right) \log \left( \frac{\Lambda
_{\perp }}{k_{+}^{\prime }-k_{+}}\right) +2a\,\,\frac{\theta \left(
k_{+}^{\prime }-k_{+}\right) }{k_{+}^{\prime }-k_{+}}  \nonumber \\
&=&2a\frac{d}{dk_{+}}\left( \theta \left( k_{+}^{\prime }-k_{+}\right) \log
\left( \frac{\Lambda _{\perp }}{k_{+}^{\prime }-k_{+}}\right) \right) .
\label{kperp}
\end{eqnarray}
We notice that there is a factor of 2 between the kernels (\ref{kspace}) and
(\ref{kperp}) for the shape function in the two regularizations. The kernel
in DR is the same as that in eq.~(\ref{kperp}), with $\Lambda _{\perp
}\rightarrow \mu $.

There is a clear analogy of the evolution of the shape function with the
Altarelli--Parisi evolution equation, but with an important difference: the
evolution kernel in this case explicitly depends on the cutoff $\Lambda
_{ET} $ of the bare theory or on the renormalization point $\mu $ if we
consider the renormalized theory\footnote{%
We thank S. Catani for a discussion on this point.}. All this is related to
the fact that the Altarelli--Parisi evolution involves a single collinear
logarithm for each loop, while our problem is double-logarithmic. Let us
discuss this point with a simple analogy. The Altarelli--Parisi evolution,
or in general the usual Callan--Symanzik evolution, is analogous to a
first-order differential equation, which is autonomous (i.e.
time-independent): 
\begin{equation}
\frac{dx}{dt}=h(x),
\end{equation}
or, in discrete form, 
\begin{equation}
x_{n+1}=O\left( x_{n}\right) ,
\end{equation}
where $O$ is a generic operator, such that the formal solution reads 
\begin{equation}
x_{n}=O^{n}\left( x_{0}\right).
\end{equation}
The evolution in eq.~(\ref{evshape}) is instead analogous to an evolution
equation of the form 
\begin{equation}
\frac{dx}{dt}=h(x,t),
\end{equation}
or, in discrete form 
\begin{equation}
x_{n+1}=O_{n}\left( x_{n}\right) .
\end{equation}
In the latter case there is a different evolution operator at each step%
\footnote{%
In double-logarithmic problems, one can obtain an autonomous differential
equation at the price of having a second-order equation, i.e. of the form 
\begin{equation}
\frac{d^{2}x}{dt^{2}}=h(x).  \nonumber
\end{equation}
This, anyway, is not an evolution equation.}.

We clarify at this point a discrepancy of a factor of 2 in the evolution
kernel K of the shape function, computed at one loop in DR in both refs. 
\cite{kg} and \cite{mn}. We agree with ref.~\cite{kg}, where the kernel is
computed from the Green function in the ET taking a $\mu $ derivative, as in
eq.~(\ref{kperp}). We disagree with ref.~\cite{mn}, where the kernel is
computed by taking the difference of the QCD Green function with the ET
Green function and then differentiating with respect to $\mu $; their kernel
is two times smaller than the one in eq.~(\ref{kperp}). The latter authors
give for the QCD amplitude, in our notation, the result 
\begin{equation}
F\left( k_{+}\right) ^{QCD}\stackrel{?}{=}\frac{1}{-k_{+}+i0}\left( -\frac{1%
}{2}\right) a\log ^{2}\left( \frac{\mu }{k_{+}-i0}\right) .
\end{equation}
They find a dependence on the renormalization point $\mu $, which we do not
find as the QCD diagram is ultraviolet - as well as infrared - finite \cite
{noi}. If we replace in their renormalization condition, which determines
the kernel, our $\mu $-independent result for the QCD Green function, 
\begin{equation}
F\left( k_{+}\right) ^{QCD}=\frac{1}{-k_{+}+i0}\left( -\frac{1}{2}\right)
a\log ^{2}\left( \frac{m_{b}}{k_{+}-i0}\right) ,  \label{nostra}
\end{equation}
we find a vanishing kernel \footnote{%
In eq. (\ref{nostra}) we have assumed $E_{X}\sim O\left( m_{b}\right) $.}.
Since the effective theory is UV-divergent and consequently $\mu $%
-dependent, we believe that there may be a problem with the renormalization
conditions. Schematically, the matrix element of a bare operator is of the
form 
\begin{equation}
\langle p|O_{B}|p\rangle =1+c\,\frac{\alpha _{B}^{dim}}{\epsilon }\left(
p^{2}\right) ^{-\epsilon }+({\rm finite\,\,for\,\,}\epsilon \rightarrow 0),
\end{equation}
where $c$ is a numerical constant, $p^{2}$ refers to an overall momentum
scale in the external state, and $\left( p^{2}\right) ^{-\epsilon }$ comes
from the one-loop integral in $D=4-2\epsilon $ dimensions; $\alpha
_{B}^{dim} $ is the bare coupling of the original $D$-dimensional theory:
for $D<4$ it has a positive mass dimension $4-D=2\epsilon $ , and it must be
kept fixed as we vary $\mu ,$ which is just an arbitrary mass scale: 
\begin{equation}
\frac{d}{d\mu }\alpha _{B}^{dim}=0.
\end{equation}
This implies the well-known condition 
\begin{equation}
\frac{d}{d\mu }\langle p|O_{B}|p\rangle =0.
\end{equation}
One usually introduces an adimensionalized bare coupling multiplying $\alpha
_{B}^{dim}$ by $\mu ^{2\epsilon },$ where $\mu $ is just an arbitrary mass
scale as we said before, 
\begin{equation}
\alpha _{B}^{{\rm adim}}\equiv \mu ^{-2\epsilon }\alpha _{B}^{dim},
\end{equation}
so that the bare Green function reads 
\begin{eqnarray}
\langle p|O_{B}|p\rangle &=&1+c\,\frac{\alpha _{B}^{{\rm adim}}}{\epsilon }%
\left( \frac{\mu ^{2}}{p^{2}}\right) ^{\epsilon }+({\rm finite\,\,for\,\,}%
\epsilon \rightarrow 0)  \nonumber \\
&=&1+c\,\frac{\alpha _{B}^{{\rm adim}}}{\epsilon }+c\,\alpha _{B}^{{\rm adim}%
}\log \frac{\mu ^{2}}{p^{2}}+\cdots  \label{whole}
\end{eqnarray}
In the minimal-dimensional scheme (MS), we include the pole term in the
renormalization constant 
\begin{equation}
Z_{MS}=1+c\,\frac{\alpha _{B}^{{\rm adim}}}{\epsilon },
\end{equation}
and the remaining terms in the matrix element of the renormalized operator, 
\begin{equation}
\langle p|O_{MS}|p\rangle =1+c\,\alpha _{B}^{{\rm adim}}\log \frac{\mu ^{2}}{%
p^{2}}+\cdots ,
\end{equation}
since $O_{B}=Z\,O_{R}$. It is only {\it after} this splitting that a
dependence on $\mu $ is introduced separately in the renormalization
constant and in the renormalized operator\footnote{%
In the notation of ref.~\cite{mn}, $\log \langle O_{B}\rangle =\partial \ln 
\widetilde{f}_{B}\left( \xi \right) /\partial \log \xi $, with $\xi \sim 1/%
\sqrt{p^{2}}.$}.

The anomalous dimension is computed from the renormalization constant
keeping $\alpha _{B}^{dim }$ fixed: 
\begin{equation}
\gamma \equiv \frac{d\log Z}{d\log \mu }=\frac{d}{d\log \mu }\left( c\,\frac{%
\alpha _{B}^{dim }\mu ^{-2\epsilon }}{\epsilon }\right) =-2c\,\alpha _{B}^{%
{\rm adim}}.
\end{equation}
It seems to us that a vanishing kernel or anomalous dimension in the
effective theory is obtained in ref.~\cite{mn} because the renormalization
constant $Z$ has been identified with the whole matrix element (\ref{whole}).

\section{Conclusions}

\label{sec_conclusion}

We have discussed the properties of decays of heavy flavour hadrons into
inclusive hadron states $X$ with an invariant mass $m_{X}$ small compared
with the energy $E_{X}$, $m_{X}\ll E_{X}$. An explicit factorization
procedure has been introduced, which holds on a integral-by-integral
basis.
It is based on:

\begin{itemize}
\item  the Cauchy theorem: it is exact and leads to the replacement of
ordinary propagators with eikonal propagators in loop integrals;

\item  the lowering of \ a hard UV cutoff \ from $\Lambda _{UV}\gtrsim m_{b} 
$ to $\Lambda _{UV}=\Lambda _{ET}\ll $ $m_{b}$, where $\Lambda _{ET}$ is the
UV cutoff of the low-energy effective theory inside which the shape
function is defined.
\end{itemize}

This technique has led us to a clean separation of all the perturbative and
non-perturbative contributions. We have found that, while the exact
kinematics of the original QCD processes involves a Wilson line {\it off}
the light-cone for the final light quark, in the low-energy effective
theory the light quark is necessarily described by a Wilson line {\it on}
the light-cone.

We have analyzed the shape function $f\left( k_{+}\right)$ 
to find out which long-distance, non-perturbative, effects are
contained in and which are not, in different regularization schemes. 
We found that $f\left(k_{+}\right)$, 
contrary to naive physical expectations, has no direct
physical meaning even in double logarithmic approximation, as it represents
a partial contribution to the complete physical process. Changing
regularization, we have explicitly shown that the leading double-logarithmic
contribution to $f\left( k_{+}\right) $ can be changed by a factor of 2,
i.e. that the shape function is substantially regularization-scheme
dependent. Only after summing the shape function with the other
contributions, is a physical, scheme-independent result recovered. We have
also shown that in lattice-like regularization the shape function factorizes
a large part of the non-perturbative effects: it contains all the double
logarithmic contributions of the full QCD process. 

Subtracting from the forward hadronic tensor $T_{\mu \nu }^{QCD}$, step by
step, each of the perturbative components, we end up with an explicit
representation of the perturbative and non-perturbative effects. For
instance, at one-loop order in DLA we have the result (see eq. (\ref
{last_step_1})): 
\begin{equation}
T_{\mu \nu }^{QCD}=\frac{s_{\mu \nu }}{2v\cdot Q}\,\frac{1}{-k_{+}+i0}\left[
1+a\,C_{h}\right] \left[ 1+a\,C_{q}\right] \left[ 1+a\,\delta Z\right] \left[
1+a\,\delta \bar{F}^{ET}\right] .  \nonumber
\end{equation}
The coefficient $C_{h}$ is a hard factor that takes into account the
fluctuations with energy $\varepsilon $ in the range $E_{X}<\varepsilon
<m_{B}.$ The other two coefficients, $C_{q}$ and $\delta Z$, are
short-distance-dominated  in lattice-like regularization schemes; $%
\delta \bar{F}^{ET}$ is long-distance-dominated in any regularization. The
tensor $W_{\mu \nu }$, i.e. the rate, is obtained (as usual) by taking the
imaginary part of $T_{\mu \nu }$. 

Another outcome of our analysis is that, contrary to single logarithmic
problems, factorization in this (double-logarithmic) problem is not exact,
even in lattice-like regularization schemes. Some long-distance effects are
present in the coefficient function: they come from gluons with a large
energy but with a very small emission angle and consequently with a small
transverse momentum. These non-perturbative effects in the coefficient
function however are expected to be suppressed on physical grounds, as they
occur in a small region of the phase space for a moderately large cutoff of
the effective theory.

Finally, we have clarified some discrepancy in the literature about the
evolution kernel for the shape-function computed in double logarithmic
approximation inside dimensional regularization.

\newpage

\begin{center}
{\rm Acknowledgements}
\end{center}

We would like to thank G. Martinelli for inspiring discussions. One of us
(U.A.) has benefited from many conversations with S. Catani. We also thank
D. Anselmi, M. Battaglia, M. Beneke, M. Cacciari, M. Ciafaloni, S.~Frixione,
M. Grazzini, M. Greco, G. Korchemsky, N. Uraltsev, B. Webber and in
particular G. Veneziano. One of us (G.R.) would like to thank the Theory
group of CERN, where this work was completed.

\newpage

\begin{figure}[b]
\par
\begin{center}
\epsfxsize=0.9\textwidth
\epsfysize=0.9\textheight
\leavevmode\epsffile{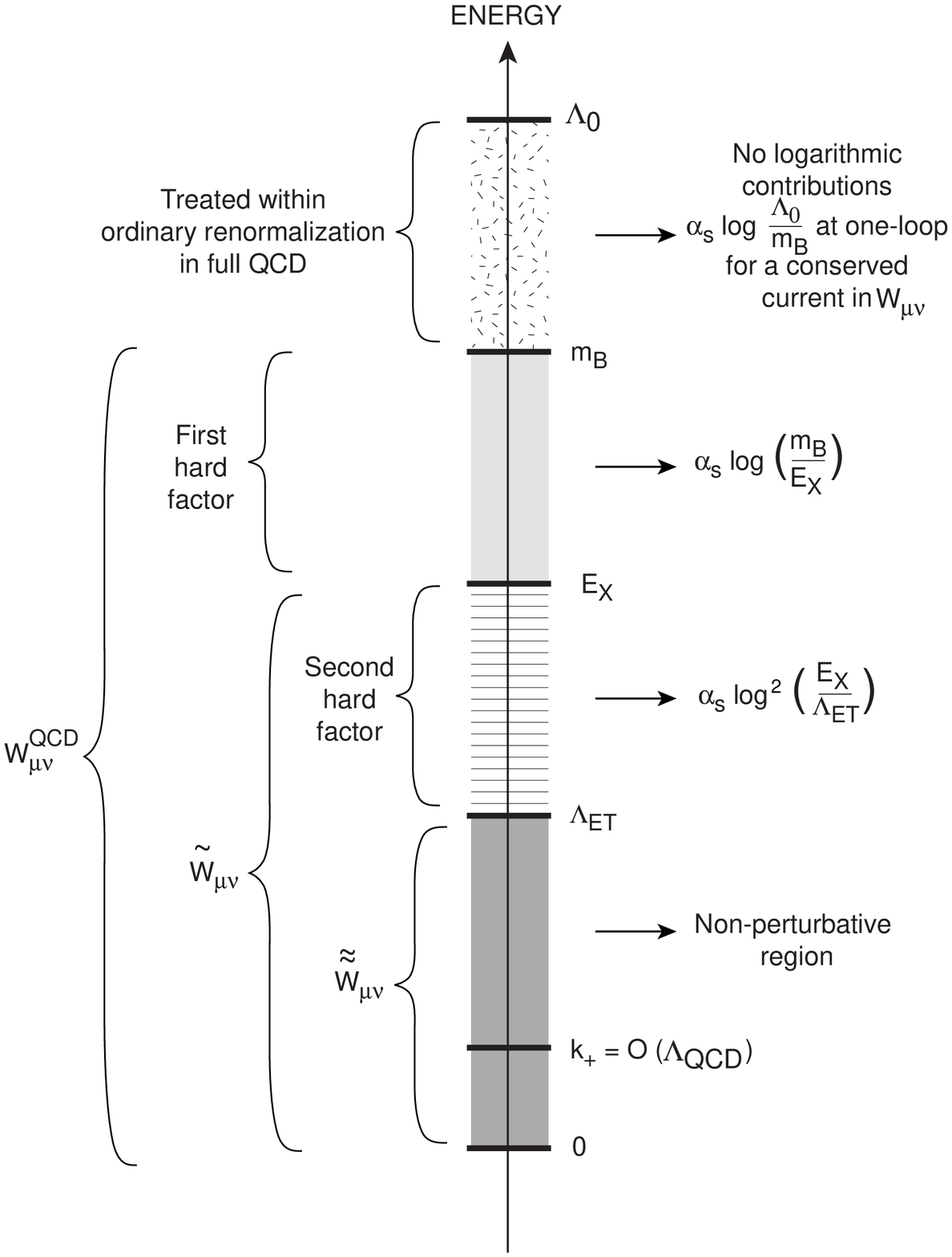}
\end{center}
\caption{{\rm {Pictorial description of factorization in the effective
theory.}}}
\label{fig:dia1}
\end{figure}

\newpage

\begin{figure}[b]
\par
\begin{center}
\leavevmode\epsffile{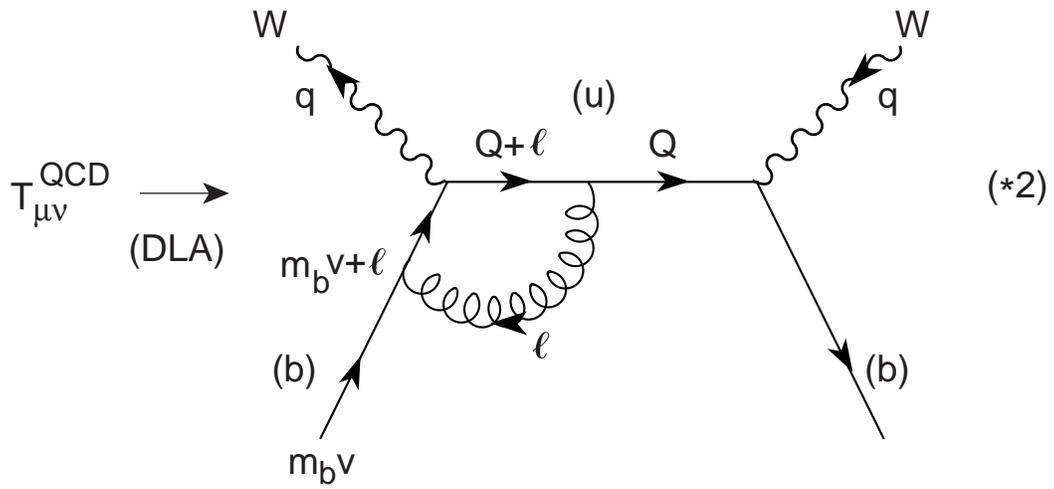}
\end{center}
\caption{{\rm {Vertex corrections to the light-cone function $F^{QCD}(k_{+})$%
}}}
\label{fig:dia2}
\end{figure}
\newpage

\begin{figure}[b]
\par
\begin{center}
\leavevmode\epsffile{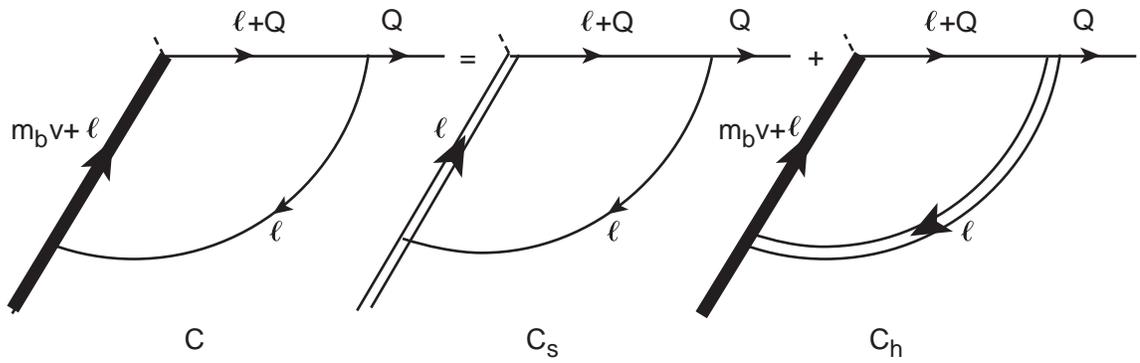}
\end{center}
\caption{{\rm {First decomposition in $T_{\protect\mu\protect\nu}$.
The thick, thin and double lines represent the massive quadratic, massless
quadratic and time-like eikonal propagators, respectively.}}}
\label{fig:dia3}
\end{figure}

\newpage

\begin{figure}[b]
\par
\begin{center}
\leavevmode\epsffile{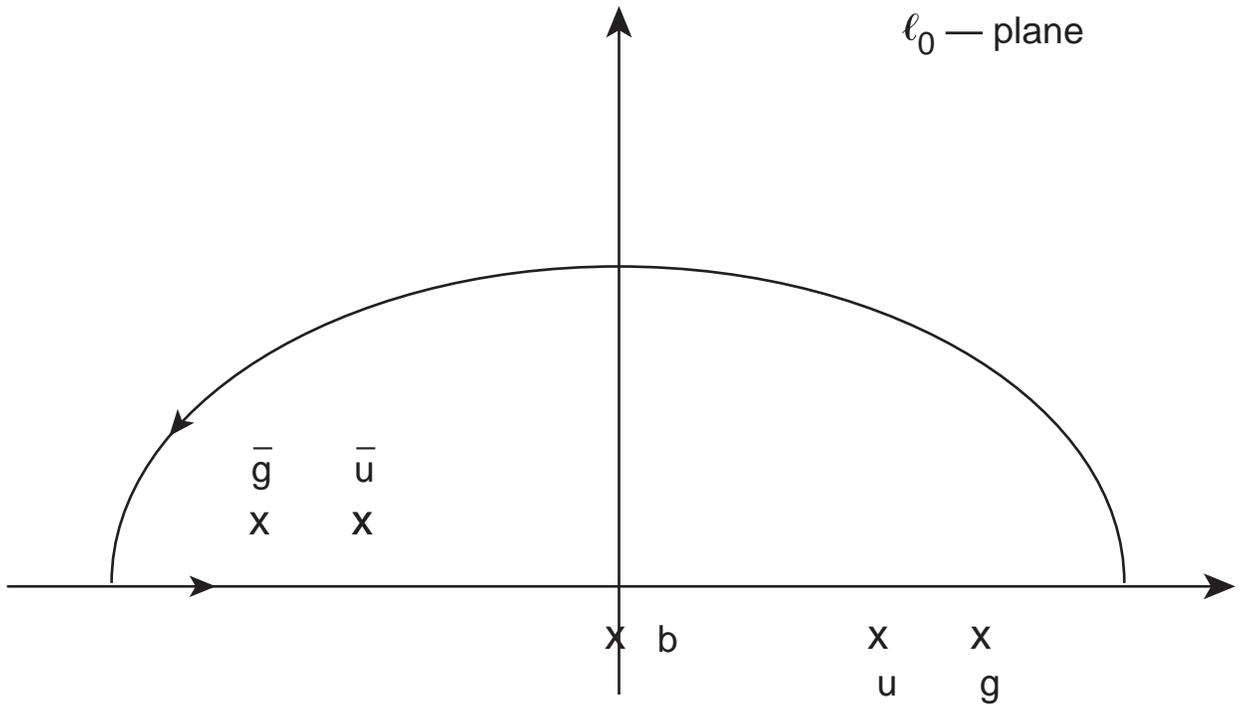}
\end{center}
\caption{{\rm {Poles of $C_s$ in the $l_0$-plane. 
The crosses labelled by $b$, $u$ and $g$ represent the particle poles in the
static beauty, $up$ and gluon propagators, respectively. The crosses
labelled by $\bar u$ and $\bar g$, instead, represent the antiparticle poles
in the $up$-quark and gluon propagators. }}}
\label{fig:dia4}
\end{figure}

\newpage

\begin{figure}[b]
\par
\begin{center}
\leavevmode\epsffile{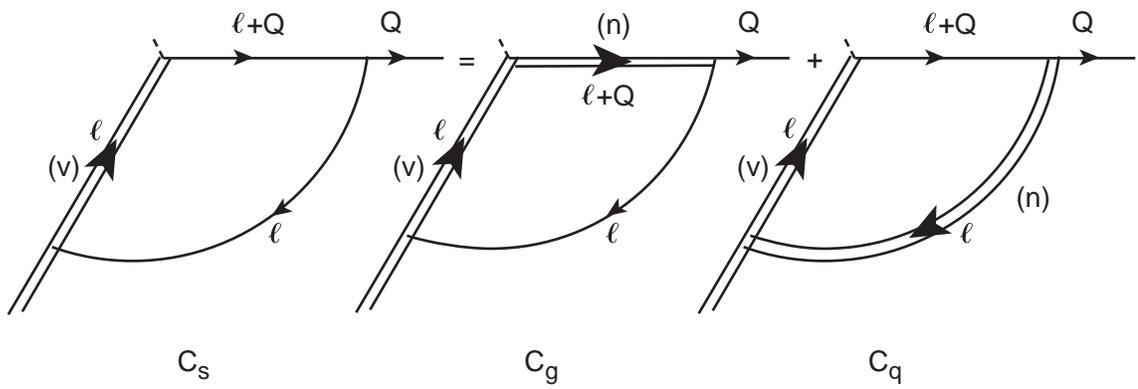}
\end{center}
\caption{{\rm {Second decomposition in $T_{\protect\mu\protect\nu}$.} 
The thin and double lines represent the massless quadratic and eikonal
propagators, respectively. The symbols (v) and (n) indicate that the eikonal
propagators must be taken along the directions $v$ and $n$. }}
\label{fig:dia5}
\end{figure}

\end{document}